\documentclass[12pt]{article}
\usepackage{amsmath}
\usepackage{graphicx,psfrag,epsf}
\usepackage{enumerate}
\usepackage{natbib}
\usepackage{url} % not crucial - just used below for the URL 

\usepackage{amsfonts}
\usepackage{amsmath}
\usepackage{amssymb}
\usepackage{graphicx}
\usepackage{booktabs}
\usepackage[width=1.25\textwidth]{caption}
\usepackage{longtable}
\usepackage{array}
\usepackage{multirow}
\usepackage{wrapfig}
\usepackage{float}
\usepackage{colortbl}
\usepackage{multirow}
\usepackage{pdflscape}
\usepackage{threeparttable}
\usepackage{threeparttablex}
\usepackage[normalem]{ulem}
\usepackage{makecell}
\usepackage{xcolor}
\usepackage{natbib}
\usepackage[nameinlink,noabbrev]{cleveref}
\usepackage{appendix}
\usepackage{algorithm}
\usepackage{algpseudocode}
\usepackage{tikz}

\def\spacingset#1{\renewcommand{\baselinestretch}%
	{#1}\small\normalsize} \spacingset{1}\def\spacingset#1{\renewcommand{\baselinestretch}%
	{#1}\small\normalsize} \spacingset{1}

\usetikzlibrary{shapes.geometric, arrows.meta, positioning, decorations.pathreplacing}

\begin{document}

\title{Fast Approximate Modular Inference for Option Price Prediction}
\author{Worapree Maneesoonthorn\thanks{%
Corresponding author; Email: Ole.Maneesoonthorn@monash.edu}, David T.
Frazier \& Gael M. Martin\thanks{%
All three authors are associated with the Department of Econometrics and
Business Statistics, Monash University. We acknowledge the important
computational work undertaken by Rub\'{e}n Loaiza-Maya at earlier stages in
this project. We thank the {Editor} and three anonymous referees 
{for providing very} valuable {guidance and }suggestions. All
three authors have been supported by Australian Research Council (ARC)
Discovery Project Grant DP200101414; Frazier has also been supported by
Discovery Early Career Researcher Award DE200101070.}}
\maketitle

\begin{abstract}
A new approximate Bayesian inferential framework is proposed that exploits
multiple information sources -- daily spot returns, high-frequency spot data
and option prices -- and enables fast calculation of probabilistic
predictions of future option prices. This approach operates directly from the theoretical option
pricing model, and does not require an explicit statistical model, or
likelihood, for the observed option prices. We demonstrate that our approach achieves predictive accuracy comparable to likelihood-based modular inference, despite not explicitly modelling option-pricing errors, while delivering substantial computational gains. Predictive accuracy based on the Heston
option pricing model is illustrated empirically for short-maturity
options, with the rapidity of real-time updates of the predictive
distributions highlighted.

\bigskip

\end{abstract}

\medskip


\spacingset{1.8} % DON'T change the spacing!
%\newpage

\section{Introduction}

Predicting future option prices is a complex task that relies on sophisticated theoretical models to capture well-documented features such as option-implied volatility smiles and smirks; see, e.g., \cite%
{bates1996jumps}, \cite{pan2002jump}, \cite{eraker2004stock}, \cite%
{fulop2015self}, and \cite{feunou2019good}. To undertake this task, modelling assumptions on the empirical mispricing of the theoretical model are typically adopted. These assumptions run the gambit from simple structures, e.g., pricing errors that are additive and homogeneous  ({\citealp{bates1996jumps,pan2002jump}}; 
{ \citealp{jiang2005model}}), to quite complex likelihood-based models (\citealp{eraker2004stock}; \citealp{lim2005parametric}; %
\citealp{martin2005implicit}; \citealp{forbes2007inference}). 
However, an inherent tension exists between
specifying a perfect theoretical model that relies on arbitrage-free
assumptions \textit{and}, simultaneously specifying a flexible statistical model that adequately
captures the deviations between the theoretical price and prices
observed in the market\spacingset{1.0}\footnote{We defer to \cite{das1999smiles}, %
\cite{jones2003dynamics}, \cite{christoffersen2008option} and \cite%
{almeida2023can} for relevant discussions.}\spacingset{1.8}. Furthermore, likelihood models that invoke additional statistical assumptions for the pricing errors often necessitate computationally intensive inferential methods, rendering predictions based on such models practically infeasible at time scales that are relevant for practitioners.

% Crucially, misspecification
% of the assumed statistical model impacts inferences on the
% theoretical pricing model, and vice versa; making assessment of the accuracy
% of either component difficult to gauge. 

In addition to option price data, daily spot price (or return) data, and
high-frequency spot data -- from which measures of volatility (\citealp{barndorff2002econometric}, \citealp{mcaleer2008realized}) or
 jumps (\citealp{huang2005relative}, %
\citealp{barndorff2006econometrics}, \citealp{maneesoonthorn2020high}) can be constructed --
are relevant for option price inference and prediction. While various combinations of these multiple information sources have been used in previous studies, most studies have exploited two sources at most (see \citealp{eraker2004stock}, \citealp{forbes2007inference}, \citealp{bollerslev2011dynamic}, \citealp{maneesoonthorn2017inference} and  \citealp{frazier2019approximate}, for examples). Indeed, to our knowledge,
only \cite{maneesoonthorn2012probabilistic} have attempted to
reconcile all three information sources and, even then, with
information in option prices captured only via the aggregated VIX index.  

In this paper we propose a fast Bayesian
approach that fully utilizes all three
information sources, while also circumventing
the need to specify statistical models for either the option pricing errors or the errors in the high-frequency measures. 
In the spirit of the likelihood-free
method of approximate Bayesian computation (ABC) (see %
\citealp{martin2024approximating}, for a recent review), inference is
conditioned on \textit{summaries }of all three sources of raw data. 
To effectively manage the multiple information
sources -- and the model structure that they induce -- we
adopt the process of modularization, or `cutting feedback', within the 
Bayesian framework (see \citealp{nott2023bayesian} for a recent review). Cutting feedback allows us to guard
against possible misspecification of certain components of the
overall model by pruning the influence of any suspect part
of the model \ -- or `module' -- on other modules that are
well-specified, while still allowing posterior
uncertainty to flow between modules (%
\citealp{chakraborty2022modularized,smith2023cutting}). Cutting
feedback has been shown to drastically improve the accuracy of
Bayesian predictions (see, e.g., \citealp{smith2023cutting} and %
\citealp{kock2026bayesian}), and our results support these findings.

Focusing on the \cite{heston1993closed} option pricing model for
ease of exposition, in Section \ref{sec:sec2} we describe the model and the main ideas underlying modular Bayesian inference in this model class. Section \ref{sec:MABI} outlines our specific
modular Bayesian approach -- which is related to, but distinct from, existing ABC methods -- and also includes a brief outline
of ABC for clarity. Extensive simulation exercises, including an assessment of alternative cutting structures,  the accuracy of the predictive distributions for option prices, and documentation of the speed with which they are computed, are presented in
Section \ref{sec:simulation}. Section \ref{sec:empirical}
outlines the empirical application of our proposed method to options written on the S\&P500
index, focusing on the one-step-ahead prediction of end-of-day market option
prices. Conclusions and discussion are provided in Section \ref{sec:conclude}. 

\spacingset{1.0}
\section{Inference for Option Pricing Using Multiple
Information Sources}\label{sec:sec2} \spacingset{1.8}

We begin by presenting the \cite{heston1993closed} option
pricing model commonly used in the option pricing literature, and on which
many more sophisticated option pricing formulae are built. In this
specific context, we proceed to outline the challenges associated
with employing multiple sources of information, motivating our particular
modular approach.\spacingset{1.0}\footnote{%
We note here that our discussion of challenges, and the treatments
thereof, can be generalized to more sophisticated option pricing \textbf{}%
models, as is discussed in Section 6.}\spacingset{1.8}

\spacingset{1.0}
\subsection{The Heston Model and Related Information
Sources}\spacingset{1.8}

The \cite{heston1993closed}
option pricing model describes the evolution of a continuous time asset
price with stochastic volatility (SV): 

\spacingset{1.0}
\begin{align}
d\ln S_{t}& =\mu dt+\sqrt{V_{t}}dW_{1t}  \notag \\
dV_{t}& =\kappa \left( \theta -V_{t}\right) dt+\sigma _{v}\sqrt{V_{t}}%
dW_{2t},  \label{eq:spot}
\end{align}%
\spacingset{1.8}where $S_{t}$ denotes the spot asset price, $V_{t}$ the latent stochastic
variance, $W_{1t}$ and $W_{2t}$ are Wiener processes and $\kappa
,\theta $ and $\sigma _{v}$ are parameters governing the dynamics of the
stochastic volatility process, with the leverage effect captured by $Corr(dW_{1t},dW_{2t})=\rho$. Prices of derivative products written on this
financial asset are formed as expectations under the corresponding
risk-neutral process. For example, the theoretical price for a European call
option with strike price $K$ and maturity $\tau $ is 
\begin{equation}
Q_{t}(k,\tau ,S_{t},V_{t},\Phi )=e^{-r\tau }E_{t}^{\mathcal{Q}}(\max
(S_{t+\tau }-K,0)),  \label{eq:Qt}
\end{equation}%
with $k=\log (K/S_{t})$ denoting the log moneyness of the option contract,
expressed relative to the spot price. This expectation $E_{t}^{\mathcal{Q}%
}(.)$, formed at time $t$ under no arbitrage conditions, is described by the
risk-neutral process $\mathcal{Q}$ 

\spacingset{1.0}
\begin{align}
d\ln S_{t}& =rdt+\sqrt{V_{t}}dW_{1t}^{\ast }  \notag \\
dV_{t}& =\kappa ^{\ast }\left( \theta ^{\ast }-V_{t}\right) dt+\sigma _{v}%
\sqrt{V_{t}}dW_{2t}^{\ast }.  \label{eq:rn}
\end{align}%
\spacingset{1.8}Here, $r$ denotes the risk-free rate, $W_{1t}^{\ast }$ and $W_{2t}^{\ast }$ are the risk-neutral
counterparts of the Wiener processes in \eqref{eq:spot}, and the
risk-neutral parameters relate to the spot parameters via $%
\kappa ^{\ast }=\kappa +\lambda $ and $\theta ^{\ast }=\kappa \theta /\kappa
^{\ast }$. The risk premium parameter $\lambda $ governs the compensation
for the stochastic nature of the volatility process. For the remainder of this paper, we assume $\rho=0$, so that in our setting, the
parameters on which \eqref{eq:Qt} depends are collected in $\Phi =(\kappa
,\theta ,\sigma _{v},\lambda )^{\top }$. 

Given the clear link between the continuous-time representation of
the SV model and both spot and derivative prices, there are three obvious
sources of information from the financial market that can be used to conduct
inference:

\begin{enumerate}
\item Observed option prices themselves, across all moneyness ($k$) 
and maturity ($\tau $) combinations traded on day $t$, collectively denoted by $O_{t}$;  
\vspace*{-0.3cm}
\item The spot daily return, computed over the trading day $t$ as $r_{t}=\ln (S_{t})-\ln (S_{t-1}),$
related to the underlying stochastic price process in (\ref%
{eq:spot}) by $r_{t}=\int_{t-1}^{t}d\ln S_{u}du$; and
\vspace*{-0.3cm}
\item Measures constructed from high-frequency spot prices,
denoted collectively by $HF_{t}$. One of the most
commonly used such measures, and the one adopted herein, %
is the jump-robust bipower variation, constructed as 
\begin{equation}
BV_{t}=\sqrt{2/\pi }\frac{M}{M-1}\sum_{i=1}^{M}|r_{t_{i}}||r_{t_{i-1}}|, \label{eq:BV}
\end{equation}%
where $r_{t_{i}}=\ln (S_{t_{i}}/S_{t_{i-1}})$, and $S_{t_{i}}$ denotes the $%
i^{th}$ spot price observed during day $t$. $BV_{t}$ is a consistent
estimator of the \textit{integrated variance (or `volatility')}, $%
\int_{t-1}^{t}V_{u}du$, which measures the variation of the continuous price
movement and is directly related to the latent variance in (\ref{eq:spot}) (%
\citealp{andersen2001distribution,barndorff2002econometric,barndorff2004power,barndorff2006econometrics}%
). 
\end{enumerate}

The stochastic price process defined in \eqref{eq:spot} produces a probability distribution for the spot return, $r_t$, via a suitable discretization. However, constructing a joint probabilistic model for all three information sources would require additional statistical assumptions for the measurement error inherent in $HF_t$ and the error that captures the\textbf{ }discrepancy between the
theoretical option price ($Q_t$) and the observed market price ($O_t$) of that option.

 In the modelling of $HF_t$, additive errors  on the logarithmic transformation of the volatility measures are typically employed; see \cite{koopman2012analysis}, \cite{maneesoonthorn2012probabilistic} and \cite{maneesoonthorn2017inference}, for examples.
With reference to option prices, given an observed market option price at time ${t}$ for an option contract with moneyness $k$ and maturity $\tau$, denoted explicitly by $O_t(k, \tau)$, the theoretical price in \eqref{eq:Qt} can be related to $O_t(k, \tau)$ via
\begin{equation}
O_{t}\left( k,\tau \right) =f(Q_{t}(k,\tau ,S_{t},V_{t},\Phi ),u_{t}),  \label{eq:Ot}
\end{equation}
where $u_{t}$ denotes the random pricing error and $f(.)$ denotes the
functional form of some statistical model. For example, in \cite{eraker2004stock}, \cite{lim2005parametric}, \cite{martin2005implicit} and \cite{forbes2007inference}, $f(\cdot)$ is assumed to be a linear function, and a range of different assumptions are adopted for $u_t$. In practice though, as noted earlier, it is challenging to both accurately specify the relation in \eqref{eq:Ot} and the theoretical pricing equation in \eqref{eq:Qt}, with misspecification of the assumed statistical model potentially impacting inferences about the theoretical pricing model, and vice versa. Moreover, irregular trading behaviours and financial market conditions may demand quite complex specifications for both $f(.)$ and $u_t$, rending realistic modelling of the link between the option price $O_{t}\left( k,\tau \right) $
and its theoretical counterpart $Q_{t}(k,\tau ,S_{t},V_{t},\Phi )$ difficult.

%\dtf{This paragraph needs work. We say that our goal is to use these datasets but also forecast option prices. Need a clearer delineation and a clearer why we want to do this.}\om{I have tried to tie the paragraphs a bit more here.}

Our goal is to propose a fast inferential method that utilizes all three sets of data to
construct predictions from the option pricing model \eqref{eq:spot}-%
\eqref{eq:rn}, without requiring an explicit statistical model either for $%
f(\cdot )$, and the distribution of $u_{t}$ in %
\eqref{eq:Ot}, or for the measurement error in $HF_t$. The predictions of future option prices produced via our method are thus not dependent on the inherently ad-hoc nature of these
statistical specifications, and the problems induced by their misspecification. 

Prior to outlining our specific methodology
in Section \ref{sec:MABI}, in the remainder of this section we 
describe how the multiple information sources \textit{could} be
combined to produce an exact (likelihood-based) Bayesian posterior --
including a modular, or cut version thereof -- thereby highlighting the
conceptual and computational issues that, in turn, motivate our proposed
approximate, or likelihood-free, modular approach.

\spacingset{1.0}
\subsection{Exact Bayesian Inference with Multiple
Information Sources}\label{sec:likelihood} 
\spacingset{1.8}
To \textit{explicitly} utilize all the
observable information in a full likelihood-based Bayesian framework%
, models are required for $r_{t}$, $HF_{t}$ and $O_{t}.$ A
discretized version of the theoretical SV\ model in \eqref{eq:spot} provides
a model for the daily return $r_{t}$. Models for $HF_{t}$ 
and $O_{t}$ entail the specification of
distributions for the noise terms that capture measurement errors and/or
microstructure noise in the case of $HF_{t}$; and in the case of $%
O_{t}$, the specification of the functional form $f(.) $\
and associated pricing error $u_{t}$ in \eqref{eq:Ot}. 

Let $\Psi _{1}$ and $\Psi _{2}$ denote the unknown
parameters on which the statistical models for $HF_{t}$ and $O_{t}$
{depend, respectively. Then, inference on all unknown structural parameters $%
\Phi =(\theta ,\kappa ,\sigma _{v},\lambda )^{\top }$, and the statistical model parameters $\Psi
=(\Psi _{1}^{\top },\Psi _{2}^{\top })^{\top }$, could proceed using
standard Bayesian methods }based on the observed data $\boldsymbol{y}%
_{t}=(r_{t},{HF}_{t},{O}_{t})^{\top }$. For notational convenience, we
denote any generic matrix of observed data as $\mathbf{X}%
_{1:T}=(X_{1},X_{2},...,X_{T})^{\top }$ and the latent variance vector as $%
\mathbf{V}_{1:T}=(V_{1},V_{2},\dots ,V_{T})^{\top }$. Given prior beliefs $%
p(\Phi ,\Psi )$, densities $p\left( \mathbf{r}_{1:T}|\mathbf{V}_{1:T},{\Phi }%
\right) $ and{\ $p\left( \mathbf{V}_{1:T}|{\Phi }\right) $ derived
from the theoretical model in \eqref{eq:spot}}, and assumed densities
for $HF_{t}$ and $O_{t}$ of the form $p\left( \mathbf{HF}_{1:T}|\mathbf{V}%
_{1:T},\Phi ,\Psi _{1}\right) $ and $p\left( \mathbf{O}_{1:T}|\mathbf{r}%
_{1:T},\mathbf{V}_{1:T},\Phi ,\Psi _{2}\right) $, respectively, inference on 
$(\Phi ^{\top },\Psi ^{\top })^{\top }$ could be conducted via the posterior 

\vspace*{-0.5cm}
\spacingset{1.0}{  
\begin{eqnarray}
p\left( \Phi ,\Psi |\mathbf{y}_{1:T}\right) &\propto &\int {p\left( \mathbf{O%
}_{1:T}|\mathbf{r}_{1:T},\mathbf{V}_{1:T},\Phi ,\Psi _{2}\right) }p\left( 
\mathbf{r}_{1:T}|\mathbf{V}_{1:T},{\Phi }\right)  \notag \\
&&p\left( \mathbf{HF}_{1:T}|\mathbf{V}_{1:T},\Phi ,\Psi _{1}\right) p\left( 
\mathbf{V}_{1:T}|{\Phi }\right) p\left( \Phi ,\Psi \right) d\mathbf{V}_{1:T}.
\label{eq:fullpost}
\end{eqnarray}
 \spacingset	{1.8}
 Using $p\left( \Phi ,\Psi |\mathbf{y%
}_{1:T}\right) $, we could then obtain a probabilistic
forecast of the future option price at time $T+h$, denoted by $O_{T+h}$, for
some known horizon $h\geq 1$, through the posterior predictive density of $%
O_{T+h}$:

\vspace*{-0.5cm}
\spacingset{1.0}{
\begin{eqnarray}
p\left( O_{T+h}|\mathbf{y}_{1:T}\right) &=&\int p\left( O_{T+h}|\mathbf{y}%
_{1:T},S_{T+h},V_{T+h},{\Phi },\Psi _{2}\right) p(S_{T+h},V_{T+h}|\Phi ,\Psi
,\mathbf{y}_{1:T})  \notag \\
&&p(\Phi ,\Psi |\mathbf{y}_{1:T})dS_{T+h}dV_{T+h}d\Phi d\Psi ,
\label{fullpred}
\end{eqnarray} \spacingset{1.8}
  where $S_{T+h}=S_{T}\exp \left( \sum_{i=1}^{h}r_{T+i}\right) $
denotes the stock price at time $T+h$. 

While the evaluation of \eqref{eq:fullpost} is
possible, albeit computationally intensive due to the complex nature in
which $\mathbf{V}_{1:T}$ enters ${p\left( \mathbf{O}_{1:T}|%
\mathbf{r}_{1:T},\mathbf{V}_{1:T},\Phi ,\Psi _{2}\right) }$, the posterior depends on the specification of the models
underpinning all three of: $p\left( \mathbf{r}_{1:T}|\mathbf{V}%
_{1:T},{\Phi }\right) , p\left( \mathbf{HF}_{1:T}|%
\mathbf{V}_{1:T},\Phi ,\Psi _{1}\right) $ and $p\left( \mathbf{O}_{1:T}|%
\mathbf{r}_{1:T},\mathbf{V}_{1:T},\Phi ,\Psi _{2}\right) $. Whilst
the model for the daily spot return is based on theoretical assumptions%
, the models for $HF_{t}$ and $O_{t}$ are both
intrinsically arbitrary. Notably, $O_{t}$ contains the observed
sequence $\{O_{t}(k,\tau ):k\in \mathbf{K}_{t},\tau \in \mathbf{\Xi }_{t}\}$%
, whose dimension can change across $t$, depending on the dimensions of the
sets of day $t$ strike prices and times to maturity, $\mathbf{K}_{t}$ and $%
\mathbf{\Xi }_{t}$ respectively, making the statistical modelling
of $O_{t}$ particularly challenging. Any
misspecification in the statistical models for $HF_{t}$ and $O_{t}$ 
would adversely affect inferences about the underlying theoretical
model, via the posterior in \eqref{eq:fullpost}, as well as affecting the
predictive density, both through the posterior $p\left( \Phi
,\Psi |\mathbf{y}_{1:T}\right) $\ itself and the model used to
specify the future option price, $p\left( O_{T+h}|\mathbf{y}%
_{1:T},S_{T+h},V_{T+h},{\Phi },\Psi _{2}\right)$ in \eqref{fullpred}. 

\spacingset{1.0}
\subsection{A Modular Approach and its
Challenges}\label{sec:twomod}\spacingset{1.8}

To reduce the possible impact of misspecification of the statistical
models for $HF_{t}$ and $O_{t}$, we \textit{could} adopt a
modular, or `cutting feedback' approach to the construction of the
likelihood. Such an approach aims to 
robustify Bayesian inference by artificially limiting the flow of
information across different components, or modules, of the assumed model (see \citealp{LiuBayarriBerger2009}). \textbf{
}To see how cutting feedback can be applied in this case, for $\Phi
_{1}=(\theta ,\kappa ,\sigma _{v})^{\top }$ and $\Phi _{2}=\lambda $, a `cut posterior' can be defined via the decomposition

\spacingset{1.0}
\begin{flalign}
p_{\mathrm{cut}}\left( \Phi,\Psi,\mathbf{V}_{1:T} |\mathbf{y}_{1:T}\right) = &\underbrace{p_{1}\left(\Phi_1,\Psi_1,\mathbf{V}_{1:T}|\mathbf{r}_{1:T},\mathbf{HF}_{1:T}\right)}_{\text{Module 1}}\times\nonumber\\&\underbrace{p_{2}\left(\Phi_2,\Psi_2|\mathbf{O}_{1:T},\mathbf{r}_{1:T},\mathbf{HF}_{1:T},\Phi_1,\Psi_1,\mathbf{V}_{1:T} \right)}_{\text{Module 2}} .
\label{eq:2modpost}
\end{flalign} \spacingset{1.8}
That is, we
could view the models for $r_{t}$ and $HF_{t}$, including the accompanying
parameters, $\Phi _{1}$ and\ $\Psi _{1}$, and latent variance, $V_{t}$, as
one module; with the model for $O_{t}$, including the parameters $\Phi _{2}$
and $\Psi _{2}$, viewed as a separate module. The implication of this
decomposition is that option prices do not inform inference about $\Phi _{1}$%
, $\Psi _{1}$ and $\mathbf{V}_{1:T}$ in Module 1, and feedback between the
two -- including the impact of misspecification of the model for $%
O_{t}$ on inference regarding $\Phi _{1}$ -- is `cut' in this sense%
 \spacingset	{1.0}\footnote{%
This particular way of `cutting feedback' is akin to the two-step estimation
often undertaken in the finance literature, whereby the spot parameters are
estimated from spot information sources, and option price parameters,
inclusive of risk premia, are \textit{calibrated} using option prices; see 
\cite{bakshi1997empirical} and \cite{fulop2015self}, for examples. However,
the Bayesian approach described here still allows for uncertainty about $%
\Phi _{1}$ and $\Psi _{1}$ to filter through\ into inference about $\Phi
_{2} $ and $\Psi _{2}$, via the structure of the second component in (\ref%
{eq:2modpost}).} \spacingset	{1.8}. 

Given (\ref{eq:2modpost}), it would then, in turn, be feasible to
construct a particular representation of the predictive using \eqref{fullpred}, but with the exact posterior $p(\Phi ,\Psi |\mathbf{y}_{1:T})$ %
replaced by its modular counterpart.
However, there are two key limitations with this approach. First, evaluation
of $p_{\mathrm{cut}}\left( \Phi ,\Psi |\mathbf{y}_{1:T}\right) $ would
require integration with respect to the full vector of latent variances, $%
\mathbf{V}_{1:T}$, which features in both modules in (\ref{eq:2modpost}).
This integration is particularly cumbersome for the (joint) conditional
density that defines Module 2, where the volatilities enter through the
complex integral defining the option price via (\ref{eq:Qt}). Second, the
decomposition of the cut posterior in \eqref{eq:2modpost} is the \textit{only} way to
conduct modular inference given the specific full likelihood
structure of the model, with the information sources relating to the spot process -- $r_t$ and $HF_t$ -- being the only sources structurally related to the parameters in $\Phi_1$, via\textbf{ \eqref{eq:spot}}. As such, this decomposition ensures that option prices
play no role whatsoever in identifying parameters for the spot price 
process, namely $\Phi _{1}$. However, option prices are information-rich, with option
contracts that are close to maturity, and with log-moneyness close to zero,
providing valuable information about the spot price process (see %
\citealp{ait2021implied}, for example). Therefore, the cut posterior
proposed in\textbf{ \eqref{eq:2modpost}} delivers more robust inference at the
substantial cost of discarding useful information for
parameter inference. 

\spacingset{1.0}
\section{The Modular Approximate Bayesian Approach} \label{sec:MABI} \spacingset{1.8}

What is required then is a way of exploiting all three types of
data, but in a more flexible manner than what can be achieved under the standard likelihood-based approach. To do so, we propose a modular Bayesian method that is `likelihood-free', and which allows for flexible feedback
cutting between different components of the assumed generating process for
the three data sources. This method is underpinned by the same basic principles of
ABC, whereby a data generating process of some sort is assumed, but is not used to structure a likelihood function based on the
observed data. Rather, the observed data enter the analysis only in the form
of summary statistics, with the analysis being `approximate' as a consequence. Before outlining the application of
modularization to this likelihood-free structure, and highlighting the
additional cut flexibility that it affords us, we first provide a brief
outline of the essence of ABC itself.

\spacingset{1.0}
\subsection{Conventional Approximate Bayesian Computation}\spacingset{1.8}

Using the notation of the current problem, and just focusing on the
structural parameters, the application of ABC would produce a posterior for %
$\Phi $ that conditions not on the full data set, $\mathbf{y}%
_{1:T} $, but on a set of summary statistics $\mathcal{S}(\mathbf{y}_{1:T})$. 
In its very simplest form, the method involves: i) simulating
artificial data from an assumed process for $\mathbf{y}_{1:T}$, 
based on repeated draws of $\Phi $ from the prior, $p(\Phi )$ (%
and draws of $\mathbf{V}_{1:T}$ from $p\left( \mathbf{V}%
_{1:T}|{\Phi }\right) $); ii) computing summary statistics from the
simulated data sets; iii) retaining the draws of $\Phi $ for which
the simulated summaries `match' the empirical summaries within a specified
tolerance, $\epsilon $; and iv) using the selected draws to produce
a posterior, $p_{\epsilon }(\Phi |\mathcal{S}(\mathbf{y}_{1:T})).$ Only as $\epsilon
\rightarrow 0$, and for $\mathcal{S}(\mathbf{y}_{1:T})$\ sufficient, is $%
p_{\epsilon }(\Phi |\mathcal{S}(\mathbf{y}_{1:T}))$ equivalent to the exact
(likelihood-based) posterior, $p(\Phi |\mathbf{y}_{1:T}).$ 
Otherwise it is a (likelihood-free) approximation only, with the quality of
the approximation being particularly dependent on the informativeness of the
summaries\textbf{ }(\citealp{marin2012approximate,FearnheadPrangle2012,pesonen2023abc}). See Algorithm \ref{alg:abc} in Appendix A for an algorithmic
representation of the steps of this simple accept/reject ABC algorithm, and %
\cite{sisson2018handbook} and \cite{martin2024approximating} for reviews of alternative versions of ABC.

The typical aim of ABC would be to approximate $p(\Phi |\mathbf{y}%
_{1:T})$ in this way when the complexity of the problem
precludes the evaluation of $p(\mathbf{y}_{1:T}|\Phi )$ but still
allows $p(\mathbf{y}_{1:T}|\Phi )$ to be simulated. The
approximate nature of the result would be the price paid for conducting
Bayesian inference in a situation in which an exact (likelihood-based)
Bayesian posterior is simply inaccessible. This is not the motivation here.
In our case, as we show below, the conditioning of posterior inference
on summaries allows us to: 1) obviate the need to provide an explicit statistical model
for $HF_{t}$ and $O_{t}$, thereby avoiding the
arbitrariness of such models; 2) have the
flexibility to leverage the different sources of information when conducting
inference about the parameters of any single model component, \textit{despite%
} the absence of statistical models for each component; and 3)
enable the production of fast predictions, despite heavy data usage.  In a similar spirit to \cite{chakraborty2022modularized}, we
achieve these three aims via a modular decomposition of a joint
posterior that: conditions on summary statistics, involves simulation from
assumed generating processes, and which serves only as an approximation to
the exact posterior, all of which are features shared by ABC. However, the
connection with ABC does not extend further than this and, as a consequence,
we refer to our method using the more generic term: modular approximate
Bayesian inference (ABI).

\spacingset{1.0}
\subsection{Modular ABI with Soft Cutting}\spacingset{1.8}

With the information content of the summaries in mind, we partition the unknown parameters in \eqref{eq:spot}-%
\eqref{eq:rn} as $\Phi =(\Phi _{1}^{\top },\Phi _{2}^{\top })^{\top }$. 
Since, by design, we do not specify statistical models for $%
HF_{t} $ and $O_{t}$, the parameters collected in $\Psi $ in %
\eqref{eq:fullpost} no longer feature in our analysis. We then partition a vector of
summary statistics $\mathcal{S}(\mathbf{y}_{1:T})$  into $\mathcal{S}_{1}(\mathbf{y}_{1:T})$ -- chosen so
that it is informative about $\Phi _{1}$ -- and $\mathcal{S}_{2}(\mathbf{y}_{1:T})$ -- chosen to 
be informative for $\Phi _{2}$. Viewed in this way, the groupings $(%
\mathcal{S}_{1},\Phi _{1})$ and $(\mathcal{S}_{2},\Phi _{2})$ define two
modules of a joint model for $\Phi $ and $\mathcal{S}(\mathbf{y}_{1:T})$
that can be used within a cutting feedback framework. Following the approach
of \cite{chakraborty2022modularized}, we operationalize this method by 
defining the cut posterior that conditions on the summaries as 
\begin{equation}
p_{\text{cut}}(\Phi |\mathcal{S}(\mathbf{y}_{1:T}))=p(\Phi
_{2}|\Phi _{1},\mathcal{S}_{1},\mathcal{S}_{2})p(\Phi _{1}|\mathcal{S}_{1}),
\label{eq:cut}
\end{equation}
where \textbf{ }$p(\Phi _{2}|\Phi _{1},\mathcal{S}_{1},\mathcal{S}_{2})$ and $p(\Phi
_{1}|\mathcal{S}_{1})$ denote the constituent posterior densities.
We employ a specific class of approximation that enables
fast computation of these constituent components; with discussion provided in
Section \ref{sec:abccutalg}. The cut posterior $p_{\text{cut}%
}(\Phi|\mathcal{S}(\mathbf{y}_{1:T}))$ can then be used to
construct the predictive distributions for future option prices, as
described in Section \ref{sec:approxpred} below.

Critically, by defining the cut posterior
as in \eqref{eq:cut}, we decouple the choice of where to cut from the underlying probabilistic model
specification. While this has come at the cost of reducing the
data information down to summary statistics, the cut posterior in %
\eqref{eq:cut} allows us to form the marginal and conditional posteriors as
we wish, creating a rich class of posteriors for inference on $\Phi$ using observed data via its summaries $\mathcal{S}(\mathbf{y}_{1:T})$.\textbf{ }%possible approximations to 
%$p(\Phi |\mathcal{S}(\mathbf{y}_{1:T}))$. 
In particular, we are no
longer bound by the modular structure dictated by the likelihood, as in %
\eqref{eq:2modpost}. 

As an illustrative example, consider the partition $\Phi
_{1}=(\kappa,\sigma _{v},\lambda)^{\top }$ and $\Phi _{2}=\theta $. 
One partition of the summaries $\mathcal{S}$ that would cut
feedback of information while still allowing for option price information to
influence inference about certain spot parameters, would be: $\mathcal{S}_{1}=(\mathcal{S}(\mathbf{r}_{1:T})^{\top },\mathcal{S}(\mathbf{O}%
_{1:T})^{\top })^{\top }$ and $\mathcal{S}_{2}=\mathcal{S}(\mathbf{HF}%
_{1:T})$. Under this particular structure, option prices would directly affect inference on two of the structural parameters, in addition to the risk premium parameter $\lambda$, something that is not possible with likelihood-based modular inference. Note, we do not necessarily advocate
this particular structure as an ideal choice, but simply 
propose it as one of many feasible constellations.

In summary, while this approach still `cuts
feedback' it does so in a much more flexible manner, and so we refer to the
approach as \textit{soft cutting}, and to any posterior so produced as a \textit{soft cut }posterior. In
Section \ref{sec:CVnum} we formally evaluate alternative cutting structures
by assessing the posterior mass concentration in a numerical setting.

\spacingset{1.0}
\subsection{Computing a Soft Cut Posterior}\label{sec:approxCut}\spacingset{1.8}
Computing the soft cut posterior requires
generating a `reference table' that
collects a large set of realizations of the parameters, generated from the prior distribution, and summary statistics, computed from the corresponding sets of simulated data. In our particular context,
we simulate artificial data from the theoretical model %
\eqref{eq:spot}-\eqref{eq:rn}, and use Gaussian mixtures to approximate
the constituent posterior components $p(\cdot |\cdot )$ in \eqref{eq:cut}. We describe each
of these steps in detail here, before providing a summary of the general
algorithm. 

\spacingset{1.0}
\subsubsection{Composition of the Reference Table} \label{sec:prior}\spacingset{1.8}

% \subsubsubsection{Choice of Priors} %\label{sec:prior}

In the context of the Heston model, we decompose the prior for $\Phi$ as
\begin{equation}
p(\Phi )=p(\sigma _{v}|\kappa ,\theta )p(\lambda |\kappa )p(\theta )p(\kappa
). \label{eq:prior}
\end{equation}
Uniform priors are employed for each component: $p(\kappa )\sim
U(a_\kappa,b_\kappa)$, $p(\theta )\sim U(a_\theta,b_\theta)$, $p(\sigma _{v}|\kappa ,\theta )\sim U(0,%
\sqrt{2\kappa \theta })$ and $p(\lambda |\kappa )\sim U(-\kappa ,0)$. The
priors respect the parameter restrictions that impose
non-negativity of the volatility process under both the empirical
and risk-neutral price paths. This prior choice induces 
uniform prior specifications over a reasonable region of the
parameter space, and is typical of that used for this type of model
in the literature (see for example, \citealp{martin2019auxiliary}).

% \subsubsubsection{Choice of Summary Statistics}

The accuracy of the cut posterior in \eqref{eq:cut} depends in large measure{\ on the
informativeness of the chosen summary statistics. As such, we} propose a set
of summary statistics associated with each source of information based on
previous research, namely, \cite{martin2019auxiliary} and \cite%
{frazier2019approximate}, in which ABC was used to estimate a\ stochastic
volatility model based on spot information, and \cite{ait2021implied}, in
which summary statistics were used in implementing a generalized method of moments (GMM) approach.

The summary statistics from the daily spot returns, %
$\mathcal{S}(\mathbf{r}_{1:T})$,
include the first four moments of the return distribution, and the score
statistic of an approximating `auxiliary' generalized autoregressive conditional heteroskedastic (GARCH) model that captures the
dynamics of the conditional returns distribution, in the spirit of %
\citet{martin2019auxiliary}. For the high-frequency measure, we utilize the bipower variation measure as defined in   
\eqref{eq:BV}, thus defining $HF_{t}=BV_{t}$. The
summary statistics, $\mathcal{S}(\mathbf{HF}_{1:T})$, comprise the first
four moments and the first-order autocorrelation of the volatility measure.
Since bipower variation is a consistent measure of integrated volatility,
its associated summaries are informative about the parameters associated
with the latent volatility process.\spacingset{1.0}\footnote{%
Note that this definition of $HF_{t}$, and the associated set of summary
statistics, may need to be modified if inference were to be conducted for
alternative SV models, most notably those that contain random price
or volatility jumps.} \spacingset{1.8}%

From the option prices, we construct summary statistics
based on the Black-Scholes option implied volatility (BSIV) surface and its
features. As proposed by \cite{ait2021implied}, the features of the BSIV
surface across the strike prices and time to maturity, provide valuable
information about the SV process in \eqref{eq:spot}, and those authors use
this information in conducting inference on the Heston SV model via GMM, in
addition to constructing the implied stochastic volatility features
nonparametrically. From the panel of closing option prices observed for each
trading period, we can extract the features of the BSIV surface, including
the level, slope and curvature in both the maturity and strike price
directions, as a set of time series over the sample period. $\mathcal{S}%
\left( \mathbf{O}_{1:T}\right) $ then comprises the first four moments and
first-order autocorrelation statistic of\ each feature of the surface, as
well as the average standardized distance between the simulated and observed
features. We remind the reader at this point that $\mathcal{S}(\mathbf{O}%
_{1:T})$ is the only source of information on the risk premium parameter $%
\lambda $. We provide more details about the construction of the option
price summaries in Appendix B.

\spacingset{1.0}
\subsubsection{A Partially\textbf{ }Misspecified Simulator} \label{sec:simulator}\spacingset{1.8}

To implement the proposed method,
we must simulate artificial data sets, which we denote by $\widetilde{\mathbf{y}}_{1:T}=(\widetilde{\mathbf{y}}_1,...,\widetilde{\mathbf{y}}_T)^\top$, with $\widetilde{\mathbf{y}}_t = (\widetilde{r}_t,\widetilde{HF}_t,\widetilde{O}_t)^\top$. 
 From each simulated data set we compute the set of summaries $\mathcal{S}(\widetilde{\mathbf{y}}_{1:T})$. For daily
spot returns, the simulated sequence, $\widetilde{\mathbf{r}}_{1:T}$, can be
constructed through a discretization of \eqref{eq:spot}, given draws of $%
\mathbf{V}_{1:T}$ and $\Phi $ from their respective prior distributions. The
vector of simulated spot prices, $\widetilde{\mathbf{S}}_{1:T}$, can be
constructed from the simulated returns, with each element of the
vector calculated as $\widetilde{S}_{t}=S_{0}\exp \left( {\sum_{i=1}^{t}%
\widetilde{r}_{i}}\right) $, with ${\widetilde{r}_{i}}$ the return at
discretization point $i.$ To bypass the need to
specify full probabilistic models for $HF_{t}$ and $O_{t}$,\ we rely on the
fact that under our theoretical model in\textbf{ \eqref{eq:spot}}-\textbf{\eqref{eq:rn} }these quantities can be directly produced.

In particular, conditional on $\mathbf{V}_{1:T}$ and $\Phi $%
, a noiseless path of high-frequency prices can be obtained using a fine
discretization of \eqref{eq:spot}, which yields, by direct calculation, a simulated sequence  $%
\{\widetilde{HF}_{t}\}_{t\ge0}$. Disregarding the possible presence of microstructure noise and discretization error, the distribution of the simulated \textbf{%
$\widetilde{HF}_{t}$ }will agree with that of\textbf{ $HF_t$}. Furthermore,\textbf{ }in the absence
of option pricing errors, the theoretical and empirical option prices
coincide, i.e., $O_{t}(k,\tau )=Q_{t}(k,\tau ,S_{t},V_{t},\Phi )$. Hence,
rather than simulating observed option prices from some
arbitrary statistical model for the pricing errors, we instead simulate
theoretical option prices $\widetilde{Q}_{t}=\{{Q}_{t}(k,\tau ,\widetilde{S}%
_{t},\widetilde{V}_{t},\Phi ):k\in \mathbf{K}_{t},\tau \in \mathbf{\Xi }%
_{t}\}$, conditional on the simulated spot prices $\widetilde{\mathbf{S}}%
_{1:T},$ plus $\mathbf{V}_{1:T}$ and $\Phi $, using \eqref{eq:Qt}.\spacingset{1.0}\footnote{Note that conditional on $\mathbf{S}_{1:T},$ $\mathbf{V}_{1:T}$ and $\Phi $, 
$\widetilde{Q}_{t}$ is a deterministic function. Singularity of the
covariance matrix of the joint summaries may be an initial concern at first
glance, with the option summaries being perfectly predictable given $\mathbf{%
S}_{1:T},$ $\mathbf{V}_{1:T}$ and $\Phi $. However, since the
summary statistics reduce the data down to vector summaries, integrating
over the latent volatilities, we can
no longer perfectly predict summaries of theoretical option prices
conditional on the parameters and the other summaries.}\spacingset{1.8}

In principle, the presence of measurement noise and discretization error for\textbf{ $HF_t$},\textbf{ }or the existence of option pricing errors for\textbf{ $O_t$},\textbf{ }implies that the simulated sequences $\{\widetilde{Q}_{t}(k,\tau ,\widetilde{S}%
_{t},\widetilde{V}_{t},\Phi ):k\in \mathbf{K}_{t},\tau \in \mathbf{\Xi }_{t}\}$ and $\{\widetilde{HF_{t}}%
\}_{t\geq 0}$ are only\textbf{ }proxies for the observed sequences $\{O_{t}(k,\tau
):k\in \mathbf{K}_{t},\tau \in \mathbf{\Xi }_{t}\}_{t\geq 0}$ and $%
\{HF_{t}\}_{t\geq 0}$, respectively.\textbf{ }However, so long as the summaries
based on $\widetilde{Q}_{t}$ and $\widetilde{HF_{t}}%
$ are informative about the unknown model
parameters $\Phi $,  the resulting inferences will not be negatively impacted.\textbf{
}Returning to the empirical option pricing equation \eqref{eq:Ot}, we see that
even if there are option pricing errors, so long as the errors $u_{t}$ do
not systematically change as the option contracts vary over $\tau ,k$, then\textbf{ }the influence of the parameters \textbf{$\Phi$ }on\textbf{ $O_t$ }is directly encapsulated by\textbf{ ${Q}_{t}$};\textbf{ }hence, the simulated\textbf{ $\widetilde{Q}_{t}$ }carries the same information about the parameters $\Phi$ as the observed prices themselves. For a specific example, if $f(\cdot )$ in \eqref{eq:Ot} is linear
in $Q_{t}$ and $u_{t}$, and if $u_{t}$ is white noise, then
summaries calculated from $\widetilde{Q}_{t}$ will, on average, be similar to
summaries calculated from the observed prices.%

Critically, the above approach delivers simulated
series $\widetilde{HF}_{t}$ and $\widetilde{Q}_{t}$ without needing to
specify full probabilistic models for $HF_{t}$ and $O_{t}$. 
Whilst this means that the `simulator
model' used within the algorithm is necessarily 
misspecified, as we have argued above this misspecified simulator,
for option prices in particular, will still deliver reliable
inferences so long as the empirical option pricing errors are not
systematic. In this way, our use of a misspecified simulator is similar
to related applications of misspecified simulators in the indirect inference
(II) literature (\citealp{GMR1993}), including the closely related examples
of misspecified simulators in measurement error models (%
\citealp{gospodinov2017simulated}), and nonparametric time series models (\citealp{frazier2021indirect}).

\spacingset{1.0}
\subsubsection{The Gaussian Mixture Approximation}\label{sec:abccutalg} 
\spacingset{1.8}
Conducting posterior inference via the soft cut distribution, $p_{\text{cut}}(\Phi
|\mathcal{S}(\mathbf{y}_{1:T}))$ defined in\textbf{ \eqref{eq:cut}} would be trivial if the joint distribution $p(\Phi ,\mathcal{S})$ were analytically tractable. However,
this is not the case even in the simplest models. Herein, we follow \cite%
{west1993approximating} and \cite{chakraborty2022modularized} and approximate the cut posterior using a Gaussian mixture model. 
Specifically, a Gaussian mixture model is fit to a large set of pseudo-data simulated from the joint distribution of $(\Phi ,\mathcal{S})$, referred to as the `reference table'.  Using the fitted joint Gaussian mixture, the\textbf{ }constituent densities, $q(\Phi _{2}|\Phi
_{1},\mathcal{S}_{1},\mathcal{S}_{2})$ and $q(\Phi _{1}|\mathcal{S}%
_{1})$, are themselves Gaussian mixture densities and can be analytically evaluated. These mixture densities then serve as an approximation to  $p(\Phi _{2}|\Phi
_{1},\mathcal{S}_{1},\mathcal{S}_{2})$ and $p(\Phi _{1}|\mathcal{S}
_{1})$, respectively. An analytical approximation to the soft cut posterior in   \eqref{eq:cut} is thereby produced as
\begin{equation}
\hat{p}_{\text{cut}}(\Phi |\mathcal{S}(\mathbf{y}_{1:T}))=q(\Phi _{2}|\Phi
_{1},\mathcal{S}_{1},\mathcal{S}_{2})q(\Phi _{1}|\mathcal{S}%
_{1}).\label{eq:phatcut}
\end{equation}
Sampling from this approximation is simple: first, sample $\Phi_1$ directly from the Gaussian mixture $q(\Phi _{1}|\mathcal{S}%
_{1})$; then, conditional on $\Phi_1$, we can sample $\Phi_2$ directly from the mixture $q(\Phi _{2}|\Phi
_{1},\mathcal{S}_{1},\mathcal{S}_{2})$, where $\mathcal{S}_1$ and $\mathcal{S}_2$ are sample statistics calculated using observed data. 

The parameters of the Gaussian
mixture model can be fitted using a standard
expectation-maximization (EM) algorithm or more recently developed 
Bayesian methods, such as those of \cite{fruhwirth2021generalized}.
We note that the Bayesian approach to fitting the Gaussian mixture
model inherently induces `posterior uncertainty in
the approximate posterior distribution'. Whilst
this has virtue, managing this additional layer of uncertainty is beyond
the scope of this work. As such, in all subsequent calculations of $\hat{p}_{\text{cut}}(\Phi |\mathcal{S}(\mathbf{y}_{1:T}))$ we use the EM algorithm built into the MATLAB
software via the `fitgmdist' function to fit the Gaussian mixture model (%
\citealp{mclachlan2000finite}). The optimal number of components, $m^{\ast }$, is selected by minimizing the Bayesian information
criterion (BIC), with the maximum number of components considered being 30.\spacingset{1.0}\footnote{As a robustness
check, we also implemented numerical experiments using
the Bayesian fitting method of \cite{fruhwirth2021generalized}. 
The resulting posterior inference and predictive results remained robust to the choice of Gaussian mixture approximation, with
results provided in Appendix D.}\spacingset{1.8}

% Once the parameters of the joint Gaussian mixture $q(\Phi ,\mathcal{S})$ have been
% estimated, the expressions for the parameters of any conditionals \marginpar{\tiny \textit{\ its conditional
% distributions What are we referencing here?? Is ambiguous. All notation
% needs to be cross-referenced.}} $q(\cdot|\cdot)$ can be computed analytically. This approach
% only requires that we are able to fit a Gaussian mixture model to the
% reference table of simulated parameters and summary statistics, which can be
% performed with off-the-shelf software reliably and efficiently, even in the
% case of the large reference tables that we face in practice. \textit{This
% final sentence seems strange given that we have already discussed
% computation in detail above. So we know what 'off-the shelf' means.}

\spacingset{1.0}
\subsubsection{The Modular ABI Algorithm}
\spacingset{1.8}
The steps needed to evaluate the soft cut
posterior are summarized in Algorithm \ref{alg:abCut}. We reiterate
that while the proposed modular ABI method is similar in spirit to
(a modularized version of) ABC, it differs from
conventional ABC in several important ways. First, we simulate the
reference table from the theoretical model in
\eqref{eq:spot}-\eqref{eq:rn} only (using discretization where needed),
expressly in order to bypass the need to specify statistical models for $%
HF_{t}$ and $O_{t}$. This means that the simulator is an incomplete
representation of the data generating process, and is misspecified in this
sense. Second, instead of obtaining the approximate posterior by matching
the distance between the observed and simulated summary statistics given an 
\textit{a priori} tolerance, as is the standard approach to ABC, we
approximate the posterior using a Gaussian mixture for the joint
distribution of the parameters and summary statistics. As well as
allowing for the cut posterior to be produced efficiently, we note that this
approach avoids the multiple matching problem encountered in
conventional ABC (see \citealp{blum2013comparative}, %
\citealp{frazier2018asymptotic}).

{  {  
\begin{algorithm} [tbph]
\caption{Modular ABI with misspecified simulator}\label{alg:abCut}
\spacingset{1.0}
\begin{algorithmic}
\State 1. Calculate the reference table: \textbf{for $i=1,2,\dots,R$}
\State \hspace{1em} a. Sample $\Phi^{(i)} \sim p(\Phi)$
\State \hspace{1em} b. Sample $\mathbf{V}_{1:T}^{(i)} \sim p(\mathbf{V}|\Phi^{(i)})$
\State \hspace{1em} c. Sample pseudo data $\widetilde{\mathbf{y}}_{1:T}^{(i)}$ using the theoretical construct as in \eqref{eq:spot}-\eqref{eq:rn}
\State \hspace{1em} d. Calculate the summary statistics $\mathcal{S}^{(i)}=\mathcal{S}(\widetilde{\mathbf{y}}_{1:T}^{(i)} )$
\State \hspace{1em}\textbf{End for loop}
\State 2. Approximate $p(\Phi,\mathcal{S}(\mathbf{y}))$ using a Gaussian mixture model, utilizing the reference table generated in Step 1.
\State 3. Sample from the approximate cut posterior components, with $\mathcal{S}_1$ and $\mathcal{S}_2$ computed from observed data $\mathbf{y}_{1:T}$:
$$\Phi_1^{(j)}\sim q(\Phi_1|\mathcal{S}_1)$$
$$\Phi_2^{(j)}\sim q(\Phi_2|\Phi_1^{(j)},\mathcal{S}_1,\mathcal{S}_2)$$
via conditional Gaussian sampling. The collective draws $(\Phi_1^{(j)},\Phi_2^{(j)})$ for $j=1,2,\dots,J$ are $J$ representative draws from $\hat{p}_{\text{cut}}(\Phi |\mathcal{S}(\mathbf{y}_{1:T}))$ in \eqref{eq:phatcut}.
\end{algorithmic}
\end{algorithm}
 \spacingset	{1.8}}}

Most importantly, approximating the posterior using
Algorithm \ref{alg:abCut} yields substantial computational benefits. The
relationship between the model parameters and the summary statistics is
learned \textit{only once} via the reference table, and any new information
that arrives in the summaries is handled via the Gaussian conditioning used
to produce the posterior. This means that when we update the posterior
approximation to produce real-time forecasts, we only require standard
Gaussian simulations, as in Step 3. of Algorithm \ref{alg:abCut}, with the summaries updated with the most up-to-date data. This feature
enables an extremely fast update of the posterior distribution and, hence,
fast updates of any required predictive distributions to be produced for
real-time forecasting.

\spacingset{1.0}
\subsection{The Approximate Predictive Option Price Distribution}\label{sec:approxpred}
\spacingset{1.8}
The exact predictive distribution of the future
option price in \eqref{fullpred} is not available under our proposed
inferential approach for three reasons. First, as we do not make any
assumptions about the probabilistic model for the observed option prices,
the first density under the integral is unavailable. Second, the lack of a
statistical model for both $HF_{t}$ and $O_{t}$ ensures that the next
density under the integral, namely the predictive $p(S_{T+h},{V}_{T+h}|%
\mathbf{y}_{1:T},\Phi )$, is also inaccessible due to the absence of the
full model required to perform the filtering needed in the construction of 
such a predictive distribution. Third, under our particular
inferential approach, we do not target the exact posterior, $p\left( \Phi |%
\mathbf{y}_{1:T}\right) .$ 

With regard to the first point, we note that the
expression for the theoretical option price remains accessible via %
\eqref{eq:Qt}, and is readily computable given $S_{t},V_{t}$ and $%
\Phi $. Consequently, we propose using the theoretical price as an input
into the predictive distribution for the observed price as follows: for $%
\mathcal{K}_{b}$ a kernel function and $b>0$  a bandwidth, an
approximation to the density for the observed option\ price at the value $%
O_{T+h}(k,\tau )$ is given by 
\begin{equation}
\mathcal{K}_{b}\left\{ {Q_{T+h}(k,\tau ,S_{T+h},V_{T+h},\Phi )}-O_{T+h}({%
k,\tau })\right\} .  %\label{kernel}
\end{equation}%
With regard to the second point, we approximate $p(S_{T+h},{V}_{T+h}|\mathbf{%
y}_{1:T},\Phi )$ by reducing the conditioning set from $(\mathbf{y}%
_{1:T},\Phi )$ to $(\mathbf{r}_{1:T},\Phi )$, resulting in $p^{a}(S_{T+h},{V}%
_{T+h}|\mathbf{r}_{1:T},\Phi )$. This approximate predictive distribution
for the future spot price and future volatilities can, in turn, be readily
constructed via the particle filter (as in \citealp{frazier2019approximate}%
), which only requires posterior draws of $\Phi $ and the likelihood
associated with the spot price (and subsequently the return) that
is derived from the theoretical model in \eqref{eq:spot}. Finally, we
replace the exact posterior, $p\left( \Phi |\mathbf{y}_{1:T}\right) $, with
the cut posterior for the structural parameters, $\hat{p}_{\mathrm{cut}}(\Phi |%
\mathcal{S}(\mathbf{y}_{1:T}))$, constructed as described in Section \ref%
{sec:approxCut}. In summary, in our setting the density of a future option
price is given by the approximate predictive distribution: 

\spacingset{1.0}
\begin{flalign}
p^{a}(O_{T+h}(k,\tau ) &\mid \mathbf{y}_{1:T})=\int \mathcal{K}_{b}\left\{ {%
Q_{T+h}(k,\tau,S_{T+h},V_{T+h},\Phi )}-O_{T+h}({k,\tau })\right\}\times
\notag \\
& p^{a}(S_{T+h},\mathbf{V}_{T+1:T+h}|\mathbf{r}_{1:T},\Phi )\times \hat{p}_{\mathrm{cut}}(\Phi |\mathcal{S}(\mathbf{y}%
_{1:T}))dS_{T+h}d\mathbf{V}_{T+1:T+h}d\Phi .
\label{eq:approxpred}
\end{flalign}\spacingset{1.8}With simulation from $\hat{p}_{\mathrm{cut}}(\Phi |\mathcal{S}(%
\mathbf{y}_{1:T}))$ produced via the Gaussian mixture sampling, the
computation of \eqref{eq:approxpred} simply relies on the evaluation of $%
p^{a}(S_{T+h},\mathbf{V}_{T+1:T+h}|\mathbf{r}_{1:T},\Phi )$, which entails
the typical computational demands of the particle filter. 

\spacingset{1.0}
\section{Numerical Analysis of Modular ABI}\label{sec:simulation} 
\spacingset{1.8}

The proposed methodology depends on the user-chosen partition of information within the `soft cut' structure, i.e., which summaries are used in the different modules within \eqref{eq:phatcut}, with the accuracy of the resultant posteriors and posterior predictives likely dependent on the specific partition. In this section, we investigate the impact of this choice on both posterior and predictive accuracy within the Heston SV model. We first compare the accuracy of both posteriors and predictives across an array of different `soft cut' structures/choices. Second, we assess the quality of the resulting predictive distributions relative to a benchmark: a modular likelihood-based posterior sampled using Markov chain Monte Carlo (MCMC), which we hereafter refer to as modular MCMC. The comparison of predictive accuracy between our approach and the modular MCMC benchmark is carried out in two settings: the case where the modular MCMC method correctly specifies the statistical models for $HF_t$ and $O_t$, and the case where it adopts misspecified models. These two scenarios allow us to directly measure the impact on predictive accuracy (or otherwise) brought about by avoiding the use of statistical models for $HF_{t}$ 
and $O_{t}$ within our procedure.

\spacingset{1.0}
\subsection{Numerical Assessment of Cut Structures}  \label{sec:CVnum} 
\spacingset{1.8}

We first explore the impact on both posterior and predictive accuracy of the chosen feedback structure in a controlled setting. To this end, we consider a simulation experiment where we can accurately measure the posterior mass around a given set of true values: we simulate $T=1000$ daily spot prices from the Heston
model with zero drift, and the SV model parameters set to $\theta =0.03$, $%
\kappa =0.05$ and $\sigma _{v}=0.03$; we assume $\rho =0$ in all numerical illustrations below. 
For each day, the theoretical option prices, assuming $%
\lambda =-0.02$, are constructed for the following specifications: $\tau \in
\{5,10,15,20,25,30,60\}$ days to maturity, and log-moneyness $k$ specified
over an unevenly spaced grid between $\pm 3\sqrt{V_{t}}\tau $, with a finer
grid used around the at-the-money options. In total, we simulate the prices
of 147 contracts per day over the 1000 trading days. The first half of the
simulated data is used to conduct posterior inference, while the latter half
is reserved\ for assessing the accuracy of the predictive distributions.  The reference table used in modular ABI is constructed using the prior decomposition in \eqref{eq:prior}, with $a_\theta=0,b_\theta=0.1,a_\kappa=0$ and $b_\kappa=0.1$. The simulation setting in which we operate allows us to calculate the posterior mass from each resultant marginal cut
posterior density over an interval centred on
the true parameter value. Specifically, we compute the
posterior mass of each chosen cut structure over the following
(marginal) regions: $\theta \in \lbrack 0.02,0.04],\kappa \in \lbrack
0.025,0.075],\sigma _{v}\in \lbrack 0.025,0.035],\lambda \in \lbrack -0.03,0]
$. 

\spacingset{1.0}
\subsubsection{Accuracy of Posterior Inference}
\spacingset{1.8}
To examine the impact of the feedback structure on the quality of the posterior inference of the structural parameters $\Phi$, we consider modular ABI procedures based on all possible two-module combinations as in \eqref{eq:phatcut}, with
the only restriction being that the risk premium parameter, $\lambda $, must
be coupled with the option price summaries%
\spacingset{1.0}\footnote{%
In this exercise, we define subsets of summaries based only on their respective information source, with all summaries within a given subset treated as a fixed unit. This enables us to assess the impact of
each data source on the posterior inference. Further subsets of summary statistics within a single information source would require investigation into the granularity of the feedback structure beyond two components, and is beyond the scope of this work.}\spacingset{1.8}. The feedback structure that returns the highest
collective posterior mass (in the fixed neighbourhood of the true parameters) is deemed to be the ``best cut structure'' for the
Heston SV model given the set of summary statistics and prior choice we
employ. With the posterior mass for each parameter representing a
probability value, the maximum possible collective mass for the four
parameters is 4, and the minimum is 0. The collective prior mass over the
evaluation region being evaluated is 1.1.

\spacingset{1.0} 
\begin{table}[tbp]
\centering
\caption{Relative posterior mass of the top 3 and bottom 3 ranked cut combinations. Combination (a) denotes
the middle-ranked cut structure; while combination (b) denotes the cut that 
coincides with two-stage estimation, with spot information used to
identify the spot parameters, and the risk premium parameter 
estimated conditional on the first stage results.}
\label{tab:Combo}
\resizebox{\textwidth}{!}{

\begin{tabular}{ccccccc}
\hline\hline
 &  &  &  &  & Relative   &Overall 80\%\\ 
Rank & $\Phi_1$ & $\mathcal{S}_1$ & $\Phi_2$& $\mathcal{S}_2$ & posterior mass & predictive coverage \\ \hline\hline
1 & $\kappa,\sigma_v,\lambda$ & $\mathcal{S}(\mathbf{O}_{1:T})$ & $\theta$ & 
$\mathcal{S}(\mathbf{r}_{1:T}),\mathcal{S}(\mathbf{HF}_{1:T})$ & 0.944 &82.0\%\\ 
2 & $\kappa,\sigma_v,\lambda$ & $\mathcal{S}(\mathbf{O}_{1:T}),\mathcal{S}(%
\mathbf{r}_{1:T})$ & $\theta$ & $\mathcal{S}(\mathbf{HF}_{1:T})$ & 0.943 &82.2\%\\ 
3 & $\kappa,\sigma_v,\lambda$ & $\mathcal{S}(\mathbf{O}_{1:T}),\mathcal{S}(%
\mathbf{HF}_{1:T})$ & $\theta$ & $\mathcal{S}(\mathbf{r}_{1:T})$ & 0.929 &82.4\%\\ 
\hline
... & ... & ... & ... & ... & ...  &...\\ 
$19^{(a)}$ & $\theta$ & $\mathcal{S}(\mathbf{HF}_{1:T})$ & $%
\kappa,\sigma_v,\lambda$ & $\mathcal{S}(\mathbf{O}_{1:T}),\mathcal{S}(%
\mathbf{r}_{1:T})$ & 0.807 &83.8\%\\ 
... & ... & ... & ... & ... & ...  &...\\ 
$37^{(b)}$ & $\theta,\kappa,\sigma_v$ & $\mathcal{S}(\mathbf{r}_{1:T}),%
\mathcal{S}(\mathbf{HF}_{1:T})$ & $\lambda$ & $\mathcal{S}(\mathbf{O}_{1:T})$
& 0.614 &77.3\%\\ 
... & ... & ... & ... & ... & ...  &...\\ \hline
40 & $\kappa,\sigma_v$ & $\mathcal{S}(\mathbf{r}_{1:T})$ & $\theta,\lambda$
& $\mathcal{S}(\mathbf{O}_{1:T}),\mathcal{S}(\mathbf{HF}_{1:T})$ & 0.474 &85.5\%\\ 
41 & $\theta,\sigma_v$ & $\mathcal{S}(\mathbf{r}_{1:T})$ & $\kappa,\lambda$
& $\mathcal{S}(\mathbf{O}_{1:T}),\mathcal{S}(\mathbf{HF}_{1:T})$ & 0.420 &77.6\%\\ 
42 & $\theta,\kappa,\sigma_v$ & $\mathcal{S}(\mathbf{r}_{1:T})$ & $\lambda$
& $\mathcal{S}(\mathbf{O}_{1:T}),\mathcal{S}(\mathbf{HF}_{1:T})$ & 0.409 &66.9\%\\ 
\hline\hline
\end{tabular}
}

\end{table}
\spacingset{1.8}

Table \ref{tab:Combo} presents the relative posterior mass\textbf{ }(column 6), calculated as the ratio of total posterior mass divided by 4 (total mass). Hence, relative posterior mass closer to one indicates tighter posteriors around the truth. We present the top and bottom three
cutting structures ranked by the relative posterior mass as
described above, along with selected special cases. All 42 cuts 
improve upon the relative prior mass of 0.275, and the posterior densities associated with the
top three cuts are plotted in Figure \ref{fig:top3}.  The maximum relative
posterior mass of 0.944 is produced by the following cut structure 
\begin{equation*}
\hat{p}_{\mathrm{cut}}(\theta ,\kappa ,\sigma _{v},\lambda |\mathcal{S}(\mathbf{y}%
_{1:T}))=q(\theta |\kappa ,\sigma _{v},\lambda ,\mathcal{S}(\mathbf{HF}%
_{1:T}),\mathcal{S}(\mathbf{r}_{1:T}),\mathcal{S}(\mathbf{O}_{1:T}))q(\kappa
,\sigma _{v},\lambda |\mathcal{S}(\mathbf{O}_{1:T})).
\end{equation*}%

We make several observations regarding the
alternative cut assessments. First, summary statistics from options, \textbf{$\mathcal{S}(\mathbf{O}_{1:T})$},\textbf{ }prove 
to be useful for posterior inference about $\kappa $ and $\sigma
_{v}$, and summaries of daily returns and the high-frequency measure, \textbf{$\mathcal{S}(\mathbf{r}_{1:T})$ }and\textbf{  $\mathcal{S}(\mathbf{HF}_{1:T})$} respectively, are informative about the unconditional variance $\theta $.
Secondly, our rankings suggest that the daily returns information $\mathcal{S}(%
\mathbf{r}_{1:T})$ adds little value to posterior inference for all parameters in any cutting structure; and tends to generate poor
inference when used on its own to construct an inferential module (as in the
bottom three cut structures). Furthermore, the sequencing of the conditioning
structure makes a difference in the quality of the concentration of the
posterior, in addition to the pairing of information sources to the
specific parameter set. For example, the second-ranked cut structure in
Table \ref{tab:Combo} uses the same pairing of information sources to the parameter subsets as the cut
ranked 19 in the same Table \ref{tab:Combo}, but with the sequence of the conditioning reversed. Yet, the second-ranked cut structure produces higher relative
posterior mass around the true parameter values. Finally, the cutting
structure that corresponds to the typical two-stage frequentist
estimation approach (\citealp{bakshi1997empirical})
performs rather poorly, returning a relative posterior mass of 0.614, 
and with a ranking of 37th out of 42.%

\begin{figure}[tbph]
\centering
\includegraphics[width=1\textwidth]{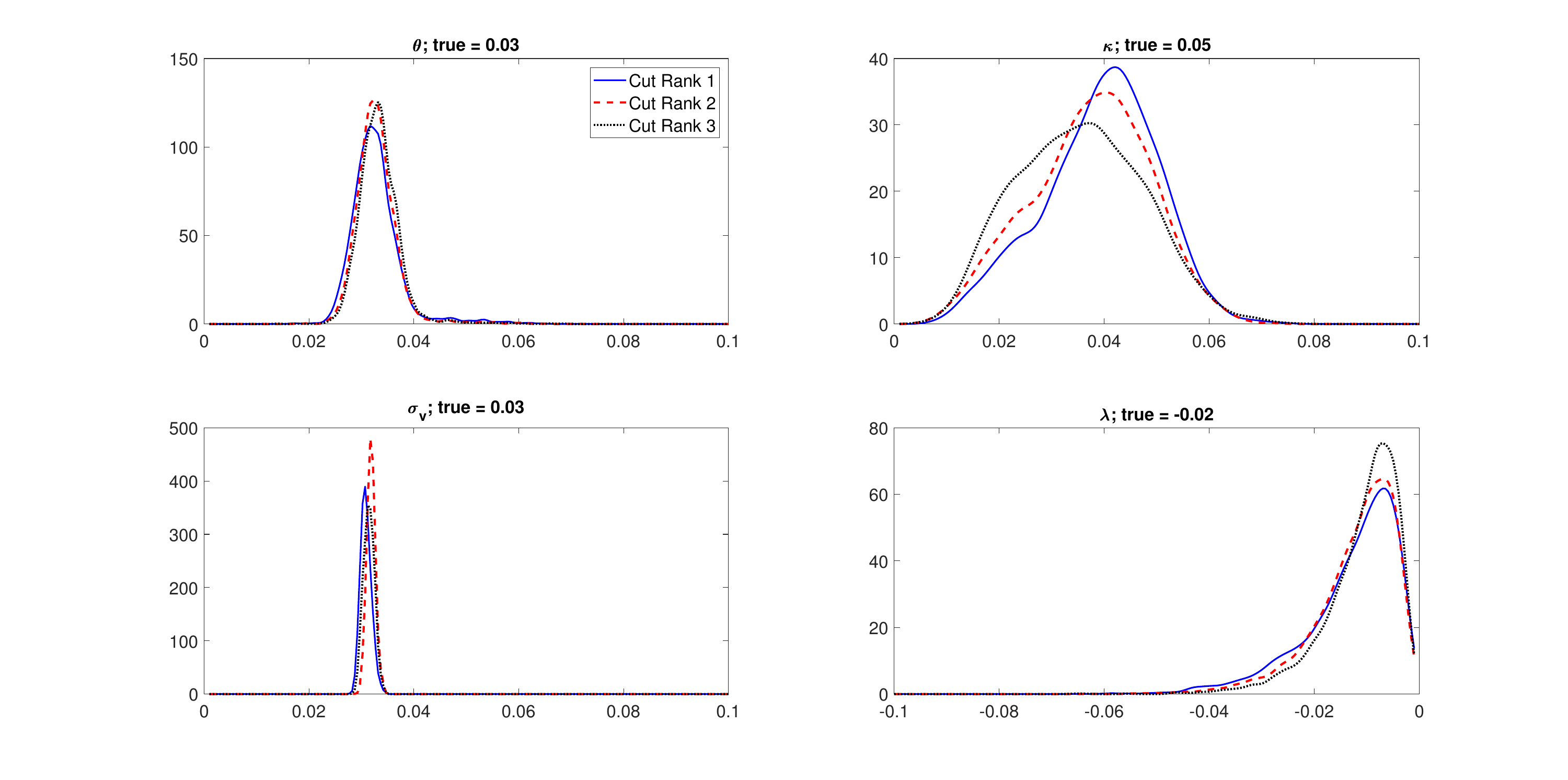}  
\caption{Comparison of the marginal posterior distributions from
the top three ranked cut structures}
\label{fig:top3}
\end{figure}
\spacingset{1.0}
\subsubsection{Predictive Accuracy}
\spacingset{1.8}
To assess the impact of different cut structures on predictive accuracy we examine the accuracy of the one-step-ahead 
predictive distributions across three specific cut structures -- the optimal cut, the rank 19 cut (middle-ranked) and the worst performing cut. We measure the predictive interval coverage of the different methods over $T_{out}=500$ trading days, and across 147
option contracts for each day. We obtain the posterior draws of the model
parameters using the first $T_{in}=500$ observations, then construct the
predictive distributions for option prices as per (\ref{eq:approxpred}),
with $h=1$, using the posterior draws from the various cut structures
documented in Table \ref{tab:Combo}. For one-step-ahead prediction, the
predictive distribution of the spot price and the latent variance, $%
p^{a}(S_{T+1},V_{T+1}|\mathbf{r}_{1:T},\Phi )$ in \eqref{eq:approxpred}, is updated using an expanding window of returns over
the out-of-sample evaluation for $T=T_{in}+i-1$ and $i=1,\dots ,T_{out}$.
Note that even though the one-step-ahead prediction of option prices only
requires the one-period-ahead filtering distribution, the evaluation of the
option price itself, via \eqref{eq:Qt}, actually requires evaluation of the expectation
of price movements for a further $\tau $ periods ahead, consistent with the
term to maturity of the option.

The final column of Table \ref{tab:Combo} summarizes the overall 80\% predictive coverages, aggregated across all option contracts and all maturities, for each corresponding cut structure. The top three cut structures, ranked by their posterior mass concentration, produce posterior predictive coverages that are close to, albeit slightly larger than, the nominal coverage. As we move down the ranking of posterior accuracy, predictive coverage deteriorates, with the bottom-ranked cut structure producing predictive distributions that can exhibit severe under-coverage. 

In Table \ref{tab:optcutcov} a more detailed breakdown of the predictive results is provided for the  top-, middle- and bottom-ranked cut structures, organized by option maturity.  These non-aggregated results reveal that for options of short maturity, there is little difference across the choice of cut: at short maturities the predictive distribution is dominated by recent information reflected in the filtering procedure in \eqref{eq:approxpred} -- an observation that is consistent with \cite{frazier2019approximate}. In contrast, at longer maturities, predictive uncertainty is driven in large part by uncertainty in the structural parameters $\Phi$. As a result, inaccurate posteriors, that is, those with less mass around the true parameter values, produce inaccurate predictions at longer maturities (see also \citealp{eraker2003impact}).

% We first compare predictive performance based on the optimal cut structure to that yielded by the
% middle-ranked structure (ranked 19 in Table \ref{tab:Combo}), with the out-of-sample
% prediction interval coverages reported in Table \ref{tab:optcutcov} below. As is evident in comparing the corresponding cells in each of the two Panels, the
% out-of-sample coverages are quite similar. We observe that both types of predictive
% provide coverage that is wider than the prescribed level (not surprising in
% the context of approximate predictions), with the lower ranked
% structure (Panel B) producing only marginally larger intervals than the optimal cut structure
% (Panel A). Thus, while there is certainly a notable difference between the
% approximate posteriors implied by the two cut structures (in terms of posterior mass concentration), there is little difference between the predictive distributions.
% This observation is in line with the earlier findings of \cite%
% {frazier2019approximate}. We also note that the difference in predictive coverages does get larger as the option maturity grows, reflecting the more dominant role of the model parameters in longer maturity options relative to the predicted latent
% volatility (see also \citealp{eraker2003impact}).

\spacingset{1.0} 
\begin{table}[tbp]
\centering \medskip {\footnotesize 
 \caption{Out-of-sample prediction interval coverage for option prices. Panel
A reports the coverages from the optimal cut structure as assessed in Section \ref{sec:CVnum}, while Panel B and C report the coverages from the middle-ranked and bottom ranked cut
structures, respectively.} \label{tab:optcutcov}
\begin{tabular}{l|ccc|ccc|ccc}
& \multicolumn{3}{c|}{\uline{Panel A: Optimal Cut}} & \multicolumn{3}{c|}{\uline{Panel B:
Rank 19 Cut}} &  \multicolumn{3}{c}{\uline{Panel C: Rank 42 Cut}} \\ 
Maturity (Days) & 80\% & 90\% & 95\% & 80\% & 90\% & 95\%  & 80\% & 90\% &95\% \\ \hline
{5} & 80.7\% & 90.9\% & 94.5\% & 80.5\% & 90.9\% & 94.7\%  & 80\% & 92.7\%&97.4\%\\ 
{10} & 81.1\% & 91.3\% & 94.9\% & 81.4\% & 91.3\% & 95.2\%  & 73.3\%& 87.6\%&95.3\%\\ 
{15} & 81.5\% & 91.6\% & 95.3\% & 82.3\% & 92.1\% & 95.6\%  & 68.3\%& 83.7\%&94.1\%\\ 
{20} & 81.7\% & 92.0\% & 95.5\% & 83.4\% & 92.9\% & 96.0\%  & 65.3\%& 81.9\%&93.8\%\\ 
{25} & 82.0\% & 92.4\% & 95.7\% & 84.2\% & 93.3\% & 96.3\%  & 63.1\%& 81.0\%&93.7\%\\ 
{30} & 82.3\% & 92.6\% & 95.9\% & 84.6\% & 93.8\% & 96.5\%  & 61.5\%& 80.5\%&94.2\%\\ 
{60} & 84.5\% & 93.6\% & 96.4\% & 87.9\% & 94.8\% & 97.1\%  & 57.1\%& 84.4\%&97.2\%\\ \hline
Overall & 82.0\% & 92.1\% & 95.5\% & 83.8\% & 92.7\% & 95.9\%  & 66.9\%& 84.5\%&95.1\%\\ \hline\hline
\end{tabular}%
}
\end{table}
%\spacingset	{1.8} 

\spacingset{1.0}
\subsection{Comparison with Modular MCMC}\label{sec:mc_noise}
\spacingset{1.8}

% With our key objective being the rapid production of
% accurate probabilistic predictions of option prices, we
% also assess our proposed method using the out-of-sample predictive
% distributions for future option prices. 

In this section, we now compare the relative performance of our modular ABI method, using the optimal cut structure, with that of a modular MCMC algorithm based on the likelihood-based cut posterior in %
\eqref{eq:2modpost}. Calculating the posterior in \eqref{eq:2modpost} requires the specification of statistical models
for $HF_{t}$ and $O_{t}$. In Scenario 1, we specify these models as:

\spacingset{1.0}
\begin{equation}
\log (BV_{t})=\log (V_{t})+\sigma_{BV}\eta _{t},  \label{eq:hf_noise}
\end{equation}
\begin{equation}
\log (O_{t}(k,\tau ))=\log (Q_{t}(k,\tau,S_{t},V_{t},\Phi
))+\sigma _{u}u_{t},  \label{eq:option_noise}
\end{equation}\spacingset{1.8}where $\eta_{t}\overset{iid}{\sim }N(0,1)$ and  $u_{t}\overset{iid}{\sim }N(0,1)$, respectively. We generate the spot data and the theoretical option prices from \eqref{eq:spot}-\eqref{eq:Qt},  using the same parameter settings described in Section \ref{sec:CVnum}. The (logarithmic) high-frequency measure $\log(BV_t)$ and the observed option price $O_t$ are then generated as in \eqref{eq:hf_noise} and \eqref{eq:option_noise}, i.e., with measurement error in $\log(BV_t)$ and pricing error in $O_t$ explicitly modelled.  Parameter values of $\sigma_{BV}=0.2$ and $\sigma_u=0.02$, respectively, are specified. We then implement modular MCMC using the noise models as specified above. That is, MCMC uses \textit{correctly specified} noise models for both $\log(BV_t)$ and $O_t$. We also note that under the DGPs employed in this subsection, the modular ABI utilizes the misspecified simulator, as described in Section \ref{sec:simulator}.\spacingset{1.0}\footnote{We also compare the modular ABI approach under the optimal cut structure to the application of a standard accept-reject ABC algorithm with full feedback, as in Algorithm \ref{alg:abc}. We found the accept-reject method to produce predictive distributions that are too wide; see Appendix C.1. Under Scenario 1, it is also possible to assess the impact of using the misspecified simulator. Critically, the use of the theoretical model as a misspecified simulator does not impinge on the accuracy of the resulting predictive distributions, with the results documented in Appendix C.2. } \spacingset{1.8}

% \gmm{The following paragraph is important, but out of place here. I reckon it needs to be positioned as a footnote somewhere below. Will require some rethink of Footnote 9.} 
% \textit{Note that the application of an MCMC algorithm to %
%  the exact (\textit{full} feedback) posterior in  \eqref{eq:fullpost}, and the production of its corresponding predictive in 
% \eqref{fullpred}, is computationally infeasible as a comparator. This is
% due to the latent variance, $V_{t}$, being embedded in the complex integral
% that defines the theoretical option price (as part of $E_{Q}(.)$ in %
% \eqref{eq:Qt}), making the posterior and predictive calculations extremely
% computationally intensive.}

In Scenario 2, we generate the $BV_t$ measure as per \eqref{eq:hf_noise}, but allow the option pricing error to take a more complex form. Specifically, we generate the observed option prices from the model 
\vspace{-0.3cm}
\begin{equation}
\log (O_{t}(k,\tau ))=\beta (k,\tau )\log (Q_{t}(k,\tau,S_{t},V_{t},\Phi ))+\sigma _{u }(k,\tau )u _{t},  \label{eq:option_complex} \vspace{-0.3cm}
\end{equation}%
with the terms $\beta (k,\tau )$ and $\sigma _{u}(k,\tau )$ controlling the
bias and the magnitude of the pricing error, respectively, over both the
moneyness and maturity of the option contracts. Equation \eqref{eq:option_complex} thus features
heterogeneity in the option pricing error, with both the bias and pricing
error magnitude increasing the further is the option contract out-of-the-money, and the further out from maturity the option contract is traded. The modular MCMC algorithm, however, still adopts the model in \eqref{eq:option_noise} (in addition to that in \eqref{eq:hf_noise}). That is,  under this scenario, the statistical model used by modular MCMC is \textit{misspecified}.

We compare our modular ABI predictive generated using %
\eqref{eq:approxpred} with the predictive obtained by \eqref{fullpred}
under the statistical model \eqref{eq:option_noise} for $O_{t}$, and %
\eqref{eq:hf_noise} for $HF_{t}$, but coupled with the modular posterior
inference outlined in Section \ref{sec:twomod}. To implement modular MCMC we apply a standard Metropolis-Hastings (MH) algorithm, allied with a particle filtering step, to obtain
Module 1 of the cut posterior defined in \eqref{eq:2modpost}, with
Module 2 obtained by an adaptive MH algorithm. This
modular MCMC benchmark took approximately 17.5 hours
to obtain 20,000 posterior draws on a Dell Precision workstation with Intel
i7 core. Comparing this with the modular ABI approach, which took a
mere twelve seconds, the computational time of our proposed algorithm was
more than four thousand times faster than that of modular MCMC.%

\spacingset	{1.0} 
% Table generated by Excel2LaTeX from sheet '3kgrid'
\begin{table}[tbph]
\caption{Out-of-sample prediction interval coverage for option prices generated from modular ABI versus modular MCMC.
Panel A reports empirical coverages under Scenario 1 - modular MCMC correctly specified; Panel B reports empirical coverages under Scenario 2 - modular MCMC misspecified.}
\label{tab:withnoise}
\centering \medskip 
\resizebox{\textwidth}{!}{
\begin{tabular}{l|ccc|ccc|ccc|ccc|}
 & \multicolumn{6}{c|} {\uline{Panel A: Scenario 1}} & \multicolumn{6}{c|} {\uline{Panel B: Scenario 2}} \\
Maturity  & \multicolumn{3}{c|}{Modular ABI} & \multicolumn{3}{c|}{Modular MCMC} & \multicolumn{3}{c|}{Modular ABI} & \multicolumn{3}{c|}{Modular MCMC} \\ 
(Days) & 80\% & 90\% & 95\% & 80\% & 90\% & 95\% & 80\% & 90\% & 95\% & 80\% & 90\% & 95\% \\ \hline
{5} & 80.6\% & 90.4\% & 95.6\% & 80.7\% & 90.6\% & 94.9\%  & 79.7\% & 90.3\% & 94.7\% & 79.8\% & 90.8\% & 94.6\% \\ 
{10} & 80.4\% & 90.8\% & 95.5\% & 81.5\% & 91.2\% & 95.1\%   & 80.6\% & 91.0\% & 95.5\% & 80.8\% & 90.9\% & 95.1\% \\ 
{15} & 80.2\% & 91.0\% & 95.4\% & 82.0\% & 91.6\% & 95.4\% & 81.9\% & 91.9\% & 96.0\% & 81.3\% & 91.4\% & 95.6\% \\ 
{20} & 80.4\% & 91.0\% & 95.5\% & 82.6\% & 92.2\% & 95.9\% & 82.6\% & 92.3\% & 96.1\% & 81.8\% & 91.6\% & 95.8\%\\ 
{25} & 80.5\% & 91.1\% & 95.6\% & 83.2\% & 92.6\% & 96.3\% & 82.7\% & 91.9\% & 96.0\% & 81.6\% & 91.6\% & 96.0\%\\ 
{30} & 81.1\% & 91.4\% & 95.8\% & 83.7\% & 92.9\% & 96.6\% & 82.1\% & 91.1\% & 95.7\% & 81.4\% & 91.2\% & 96.0\% \\ 
{60} & 83.7\% & 92.5\% & 96.8\% & 86.5\% & 94.5\% & 97.7\% & 67.7\% & 79.3\% & 86.3\% & 69.6\% & 82.1\% & 89.7\% \\ \hline
Overall & 81.0\% & 91.2\% & 95.7\% & 82.9\% & 92.2\% & 96.0\% & 79.6\% & 89.7\% & 94.3\% & 79.5\% & 89.9\% & 94.7\%\\ \hline\hline

% \multicolumn{7}{c} {}  \\
% \multicolumn{7}{c} {\uline{Panel B: Scenario 2 - Modular MCMC misspecified}} \\
% & \multicolumn{3}{c|}{Modular ABI} & \multicolumn{3}{c}{
% Modular MCMC} \\ 
% Maturity & 80\% & 90\% & 95\% & 80\% & 90\% & 95\% \\ \hline
% {5} & 79.7\% & 90.3\% & 94.7\% & 79.8\% & 90.8\% & 94.6\% \\ 
% {10} & 80.6\% & 91.0\% & 95.5\% & 80.8\% & 90.9\% & 95.1\% \\ 
% {15} & 81.9\% & 91.9\% & 96.0\% & 81.3\% & 91.4\% & 95.6\% \\ 
% {20} & 82.6\% & 92.3\% & 96.1\% & 81.8\% & 91.6\% & 95.8\% \\ 
% {25} & 82.7\% & 91.9\% & 96.0\% & 81.6\% & 91.6\% & 96.0\% \\ 
% {30} & 82.1\% & 91.1\% & 95.7\% & 81.4\% & 91.2\% & 96.0\% \\ 
% {60} & 67.7\% & 79.3\% & 86.3\% & 69.6\% & 82.1\% & 89.7\% \\ \hline
% Overall & 79.6\% & 89.7\% & 94.3\% & 79.5\% & 89.9\% & 94.7\% \\ \hline\hline
\end{tabular}%
}
\end{table}
 \spacingset	{1.8}

Table \ref{tab:withnoise} reports the predictive coverage under Scenario 1 (modular MCMC correctly specified) in Panel A, and under Scenario 2 (modular MCMC misspecified) in Panel B. As can be seen in Panel A, even though we produce prediction intervals using only a
conditional distribution for the \textit{theoretical }option price in the
case of modular ABI (as per (\ref{eq:approxpred})), the interval
coverages remain quite accurate in the presence of both high-frequency
microstructure noise and pricing errors in the option prices. We note that the correctly specified modular MCMC algorithm produces coverages that are in line with our method, also producing predictive
intervals that are slightly too wide. However, the ABI method
remains slightly more accurate overall, and in particular for the longer
maturity options. 

In Panel B it can be seen that the modular ABI and misspecified modular MCMC
methods both produce empirical coverages that are close to the nominal
levels for short-maturity options, where the bias term and option pricing error
magnitude are both small. As the bias and pricing error magnitude get larger with longer-maturity options, both methods produce less accurate prediction intervals.
When the bias and pricing error magnitude are the largest at 60 days
maturity, both methods produce prediction intervals that are too narrow.

In summary, at least in this experiment, we observe no meaningful differences between the empirical coverages obtained from the proposed modular ABI approach and modular MCMC, despite the fact that modular ABI produces predictive option prices directly from the theoretical option pricing formula. Hence, in a practical setting where fast updates are required, modular ABI is strictly preferable due to its \textit{vastly} superior computational efficiency, achieving - we reiterate -  a \textit{four-thousand-fold} reduction in computational time relative to modular MCMC. We note that the application of an MCMC algorithm to the \textit{exact} (full feedback) posterior in  \eqref{eq:fullpost}, and the production of its corresponding predictive in \eqref{fullpred}, is simply computationally infeasible as a comparator.\spacingset	{1.0}\footnote{ Due to the latent variance, $V_{t}$, being embedded in the complex integral
that defines the theoretical option price (as part of $E_{Q}(.)$ in %
\eqref{eq:Qt}), one single MCMC draw obtained from the full
feedback exact posterior takes approximately half an hour to
produce. Extrapolation of this estimate implies that obtaining the equivalent of 20,000 posterior draws
from the full posterior would take over a year (specifically, $(30\times20,000/60)/24\approx$ 416 days) on the same computational facility used for modular ABI.%
} \spacingset	{1.8} 

\spacingset{1.0}
\section{ Empirical Illustration and Forecasting Implications} \label{sec:empirical} 
\spacingset{1.8}
In this section we provide an empirical illustration
of our proposed method in predicting the market prices of option contracts
written on the S\&P500 index. We employ the \cite{heston1993closed}
theoretical option pricing model for this illustration, with full
acknowledgement that this theoretical option pricing model is most likely
misspecified in the real data setting. We highlight here that the purpose of
our work is not to search for the best theoretical pricing model. Rather, we
use the empirical results to illustrate that the method proposed in the
paper allows for reasonably accurate predictive distributions of option
prices to be produced, and in rapid time, despite the certain
misspecification of the theoretical model. In particular, we highlight the
speed with which posterior updates of the model parameters are produced.
This feature is critically important in the context of producing timely and
accurate predictions of option prices when the market conditions are
constantly changing and/or the theoretical pricing model does not capture
all characteristics of the data. We emphasize that \textit{both}
modular and exact MCMC are computationally infeasible in the 
rolling-window updating scheme that we employ here. 

\spacingset{1.0}
\subsection{ Data Description}
\spacingset{1.8}
Our empirical analysis utilizes all three data types as
they relate to the S\&P500 market index, with the daily returns data,
high-frequency measures of volatility and market option prices extracted for
the trading days between 1 January 2004 and 31 December 2021. We obtained
the intraday observations on the index from Refinitiv DataScope, and
constructed bipower variation for day $t$ ($BV_{t}$) using five-minute
intraday intervals. The return on day $t$ ($r_{t}$) is computed as the log
price difference between the daily closing and opening prices. 

Daily observations on the market closing prices of
European call option contracts written on the S\&P500 market index were
obtained from the Chicago Board Options Exchange (CBOE) DataShop. In our
analysis below, we consider option contracts that are six months or less in
maturity, with strike prices that are within $\pm 20$\% of the spot market
index. We exclude contracts that have trade volume less than 10 on a daily
basis. We reserve the data of the first 500 trading days between 2004 and 2006
as our initial estimation window. In total, we construct out-of-sample
predictions of 496,887\ closing market option prices across the 4,528
trading days. On each trading day, the number of active contracts may vary,
with the maturity ($\tau $) and moneyness ($k$) also varying over time. We
note, in particular, that the derivative market has become much more active
in the past decade, with the number of active call option contracts rising
from an average of 50 active contracts per day in the early 2000s, to
approximately 200 active contracts per day by the end of 2021.

\spacingset{1.0}
\subsection{Real-Time Prediction of
Option Prices}\spacingset{1.8}

In the forecasting exercise, we adopt a rolling-window
scheme to update the Heston model parameters and to calculate the
one-day-ahead predictive distributions of option prices for the next trading
day, with $h=1$, using the daily closing call option prices across the
strikes and maturities described earlier. The rolling-window estimation by
modular ABI using the Gaussian mixture model is\ -- as in the simulation
exercise -- extremely fast to undertake, with each posterior update\ taking
just under twelve seconds on a Dell Precision workstation with an Intel i7
core. We employ a fixed rolling window of $500$ trading days, so that the
predictive distribution, constructed as per \eqref{eq:approxpred}, is
updated daily with $h=1$ and $T=500$. The soft cut posterior
distribution is constructed using the optimal cut structure as per our assessment of alternative cut structures in
Section \ref{sec:CVnum}. In addition, the cut posterior accounts for
changing risk-free rates over the sample\spacingset{1.0}%
\footnote{%
We construct the Gaussian mixture based on preliminary simulations over a
grid of values for the risk free rate. The average Federal Funds rate over
the fixed rolling window is then used in the calculation of the cut
posterior. The prior described in Section \ref{sec:prior} is
used in all our posterior inference in this section.}\spacingset{1.8}.

 Figure \ref{fig:fixedWindow} summarizes the evolution
of the (cut) posterior median for each parameter, over the rolling-window
samples. Reiterating that the Heston pricing model is almost certainly
misspecified, we observe distinct changes in the structural parameters of
the model over time. The unconditional variance, $\theta $, and the
volatility of volatility, $\sigma _{v}$, both increase with the level of
market volatility, peaking during the midst of the Global Financial Crisis
between 2008 and 2009, and more recently during the COVID-19 pandemic. We
also note that the persistence parameter $\kappa $ is relatively high
compared to previous studies, most likely due to the lack of incorporation
of jumps in the pricing process. The risk premium parameter $\lambda $ also
changes overtime, indicating the changing compensation for random volatility
required by the option market.

\spacingset{1.0}
\begin{figure}[tbh]
\centering
\includegraphics[width=1.0\textwidth]{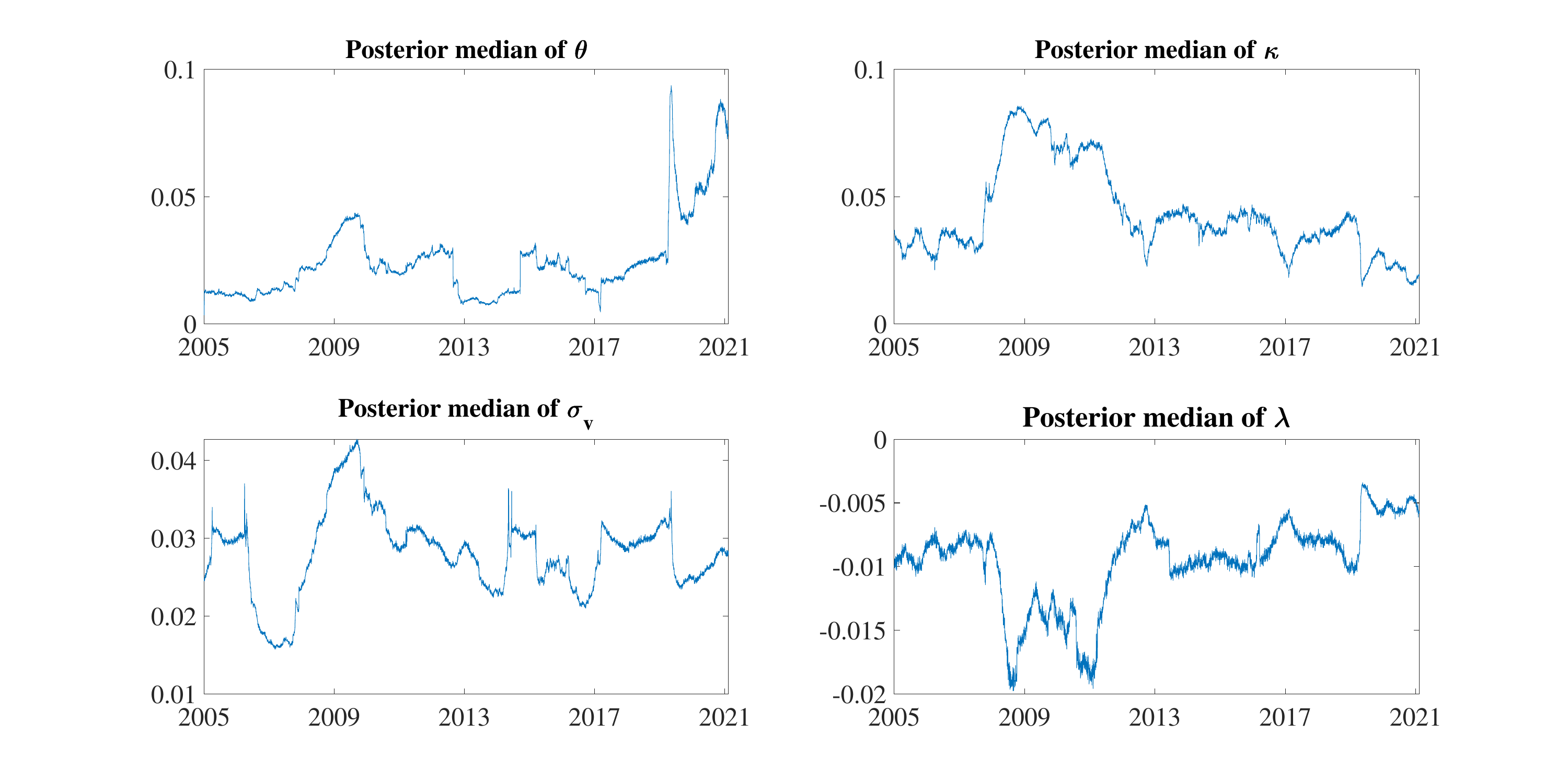}  
\caption{Posterior median of the Heston model parameters, using the fixed
rolling-window estimation scheme. The risk-free interest rate over the
estimation period is taken to be the average Federal Funds rate over the
estimation window.}
\label{fig:fixedWindow}
\end{figure}
 \spacingset	{1.8}

The predictive results are evaluated over the period
from 3 January 2006 to 31 December 2021. Table \ref{tab:pred_opt} documents
the empirical coverage of the prediction intervals, for a selection of option maturities, with more detailed results provided in
Appendix E.2. The overall coverage statistics, reported in the bottom row of
the table, reveal that the Heston option pricing model coupled with
modular ABI generates predictive option price distributions that are too
narrow overall. That said, we observe that predictions for short-maturity
options are associated with very reasonable coverage statistics,
particularly for options that mature within one calendar week. The method is
most accurate in predicting options that mature within a trading day, with
the empirical coverage statistics hovering just under the nominal coverage
levels. We also investigated the empirical coverage statistics for the 90\%
prediction interval categorised by the moneyness ($k$) of the option
contracts, with the results documented in Appendix E.2. In summary, the
prediction interval produces more accurate empirical coverage if the option
strike price is within close proximity of the closing spot price, that is,
when $|k|$ is close to zero. This observation is not surprising, given that
the \cite{heston1993closed} option pricing model is likely misspecified in 
terms of capturing the extreme tails that impact the
prices of options with extreme moneyness.

% Table generated by Excel2LaTeX from sheet 'Options'
\begin{table}[htbp]
\caption{Empirical coverage of the predictive distribution of option prices
by maturity.}
\label{tab:pred_opt}  \centering \medskip %
 \spacingset	{1.0}{\footnotesize 
\begin{tabular}{rrrrr}
\multicolumn{1}{l}{Maturity} & {95\%} & {90\%} & {80\%} & \multicolumn{1}{l}{
\# of contracts} \\ \hline
1 day & 93.6\% & 89.9\% & 80.0\% & 3681 \\ \hline
5 days & 91.4\% & 85.8\% & 76.9\% & 3144 \\ 
2 weeks & 86.5\% & 79.0\% & 67.5\% & 38458 \\ \hline
1 month & 82.9\% & 75.2\% & 63.6\% & 176808 \\ 
3 months & 65.6\% & 56.9\% & 45.8\% & 86971 \\ 
6 months & 61.4\% & 52.9\% & 42.2\% & 16569 \\ \hline
Overall & 73.2\% & 64.7\% & 53.3\% & 496887 \\ \hline\hline
\end{tabular}
}
\end{table} \spacingset{1.8}
In summary, despite the fact that we do not cater
for any pricing errors in constructing the approximate predictive
distribution of option prices, and that the predictive is based on the
Heston model, which is well known to be misspecified as a theoretical
pricing model, our method is still capable of producing reasonable
predictive coverages. We purport once again, that the fact the model
parameters are able to be updated daily to reflect the most recent changes
in the market environment -- due to the extreme speed of the modular ABI
method -- is critical to this outcome.

\spacingset{1.0}
\section{Discussion and Conclusions} \label{sec:conclude} 
\spacingset{1.8}
We propose a computationally
efficient method to predict option prices, reconciling three key information
sources in the process. Using an approximate Bayesian inference 
(ABI) framework, we use the theoretical option pricing model as a
misspecified simulator in constructing the summary statistics, thereby
avoiding the specification of arbitrary statistical models for both option
pricing errors and the microstructure noise that characterizes
high-frequency measures. With modular ABI relying only on summary
statistics, modularization of the posterior is not dictated by the rigid
structure of the joint likelihood. The modular posterior, in turn
approximated using Gaussian mixture approximations, is extremely fast to
compute, with real-time updates of this posterior, and the subsequent
 predictives, completely feasible in practice. In a simulation context, we show that our method produces option price predictions with accuracy on a par with a modular MCMC algorithm, whether the latter is correctly specified or not, in addition to being orders of magnitude faster. In an
empirical setting, despite the fact that the Heston model is almost
certainly misspecified as a theoretical option pricing model, the predictive
distribution produced by our method still produces very reasonable coverage
for short-maturity contracts, due to the continuous real-time updates of the
cut posteriors that feed into the predictions. 

Even though our description of the inferential
framework and the simulation and empirical analyses conducted are limited to
the Heston model, the proposed framework is applicable to other theoretical
models that incorporate additional stochastic features. The
framework accommodates all cases in which the option prices can be
evaluated, either by numerical or simulation methods. It does not require 
closed-form solutions for the option prices, nor any
direct mapping of the model parameters to the summary statistics. In
the case of more sophisticated models, attention would need to be paid 
to the construction of summary statistics 
that are informative about the additional structural parameters, and
careful consideration given to the form of modularization used. For example,
an option pricing model that includes price jumps may require additional
high-frequency measures of jump timing and/or size and/or specific
option price summaries that are related to extreme moneyness options, with
the treatment of the associated jump parameters in any
modular structure coupled with these additional summary measures. 
In particular, extension of the modular ABI approach in Section \ref%
{sec:MABI} to a multi-module setup may be required, with such a %
mechanism relying on standard conditional probability theory. The timely
updating scheme, along with the characterization of uncertainty in
future option prices, will also likely contribute to further advances in 
the formation of effective derivative trading rules if 
linked with appropriate consideration of investor behaviour. We
leave such considerations for future research. 

\spacingset{0.8}{\footnotesize 
\bibliographystyle{Myapalike}
\bibliography{abccut.bib}}

\newpage \pagenumbering{arabic} %
\setcounter{page}{1} 

\spacingset{1.8}

\begin{center}
{  {  \ {\large \textbf{SUPPLEMENTARY APPENDIX for
``Fast Approximate Modular Inference for Option Price Prediction''}} }}
\end{center}

\section*{ Appendix A: Conventional ABC
Algorithm}

The simplest approach to applying ABC is via the
accept-reject sampling described in Algorithm \ref{alg:abc}. Under this scheme, we sample the parameters
and the latent variances from the priors. Conditional on the $%
i^{th} $ prior draws, ${\Phi }^{(i)}$ and $\mathbf{V}_{1:T}^{(i)}$, the 
pseudo-data $\widetilde{\mathbf{y}}_{1:T}^{(i)}$ is then sampled
from the assumed model. The collection of ${\Phi }^{(i)}$ and the
corresponding summary statistics $\mathcal{S}(\widetilde{\mathbf{y}}%
_{1:T}^{(i)})$ serves as the reference table that captures the
relationship, \textit{a-priori}, between the model parameters and the
summary statistics used in inference. Draws of the parameters that yield 
pseudo-data that are deemed to be close to the observed data,
measured by the distance between the simulated and observed summary
statistics, are accepted as posterior draws. Here, $\epsilon $ denotes a
(small) tolerance level.

\setcounter{algorithm}{0} \renewcommand{%
\thealgorithm}{A\arabic{algorithm}}  \spacingset	{1.0}%
\begin{algorithm} [H]
\caption{ABC: Accept-Reject}\label{alg:abc}
\begin{algorithmic}
\State 1. Calculate the reference table. \textbf{For $i=1,2,\dots,R$}
\State  \hspace{1em} a. Sample $\Phi^{(i)} \sim p(\Phi)$
\State  \hspace{1em} b. Sample $\mathbf{V}_{1:T}^{(i)} \sim p(\mathbf{V}_{1:T}|\Phi^{(i)})$
\State  \hspace{1em} c. Sample pseudo-data $\widetilde{\mathbf{y}}_{1:T}^{(i)} \sim p(\mathbf{y}|\mathbf{V}_{1:T}^{(i)},\Phi^{(i)})$
\State  \hspace{1em} d. Calculate the summary statistics $\mathcal{S}(\widetilde{\mathbf{y}}^{(i)}_{1:T} )$
\State  2. Select $\Phi^{(i)}$ such that 
$$
d\left\{\mathcal{S}(\mathbf{y}_{1:T}),\mathcal{S}(\widetilde{\mathbf{y}}_{1:T})\right\} \leq \epsilon
$$
\end{algorithmic}
\end{algorithm} \spacingset	{1.8}

\section*{Appendix B: Summary
Statistics from Option Prices}

\setcounter{equation}{0} \renewcommand{\theequation}{B%
\arabic{equation}}
We employ the Black-Scholes option-implied volatility
(BSIV), calculated across different maturities ($\tau $) and log moneyness ($%
k$). The BSIV, one of many \textquotedblleft Greeks\textquotedblright\
reported alongside the market option prices themselves, is defined as 
\begin{equation}
\Sigma _{t}\left( \tau ,k\right) =P_{BS}^{-1}\left( \tau ,k,O_{t}\left(
k,\tau \right) \right) ,  \label{eq:bsiv}
\end{equation}%
where $P_{BS}\left( .\right) $ denotes the Black-Scholes pricing formula.
The BSIV is often used to gauge how well alternative pricing models mimic
the market-implied BSIV patterns, with option-implied volatility `smiles'
and `smirks' signalling the mis-pricing of the Black and Scholes option
pricing model. More recently, \cite{ait2021implied} and \cite{ait2021closed}
derive the relationship between the characteristics of the BSIV and the
underlying pricing process, and exploit this relationship in conducting GMM
inference about the assumed process. 

Here, we utilize summary statistics from the BSIV\
surface in the spirit of \citet{ait2021implied}. As these authors
demonstrate, the shape characteristics of the surface can be computed using
the following expression: 
\begin{equation}
\Sigma ^{\left( J,\mathbf{L}\left( J\right) \right) }\left( \tau ,k\right)
=\sum_{j=0}^{J}\sum_{i=0}^{L_{j}}\sigma _{t}^{\left( i,j\right) }\tau
^{i}k^{j},  \label{exp}
\end{equation}%
where $\Sigma ^{\left( J,\mathbf{L}\left( J\right) \right) }$ is the Taylor
series expansion of the BSIV defined in (\ref{eq:bsiv}), constructed around
the term to maturity, $\tau $, and moneyness, $k$, of order $J$ and $\mathbf{%
L}\left( J\right) $, respectively. Here, $\mathbf{L}\left( J\right) =\left(
L_{0},L_{1},...,L_{J}\right) $, with $L_{j}\geq 0$. Based on the closed-form
expansion in (\ref{exp}), we can extract the daily quantities of $\sigma
_{t}^{\left( i,j\right) }$ by means of a linear regression 
\begin{equation}
\Sigma ^{data}\left( \tau _{t}^{\left( m\right) },k_{t}^{\left( m\right)
}\right) =\sum_{j=0}^{J}\sum_{i=0}^{L_{j}}\beta _{t}^{\left( i,j\right)
}\left( \tau _{t}^{\left( m\right) }\right) ^{i}\left( k_{t}^{\left(
m\right) }\right) ^{j}+\epsilon _{l}^{\left( m\right) }\ \ \text{\ for }%
m=1,2,...,M_{t},  \label{sigij_reg}
\end{equation}%
with $M_{t}$ denoting the number of option contracts observed at time point $%
t$. The quantities $\sigma _{t}^{\left( i,j\right) }$ for a particular
period $t$ are easily estimated as 
\begin{equation}
\widehat{\sigma }_{t}^{\left( i,j\right) }=\widehat{\beta }_{t}^{\left(
i,j\right) },  \label{sigij}
\end{equation}%
where the $\widehat{\beta }_{t}^{\left( i,j\right) }$ are the ordinary least
squares estimates of $\beta _{t}^{\left( i,j\right) }$ in (\ref{sigij_reg}). 

We use the same expansion, and the resultant estimates
of $\sigma _{t}^{\left( i,j\right) }$ given in (\ref{sigij}), to construct
summary statistics in our approximation to conduct inference about the SV
model. Unlike the GMM approach, where a parametric function relating the $%
\widehat{\sigma }_{t}^{\left( i,j\right) }$ to the structural parameters of
the model is required to produce estimates of the unknown parameters, the
ABC approach simply uses the observed $\widehat{\sigma }_{t}^{\left(
i,j\right) }$ as data from which summary statistics are constructed. The
summaries $\mathcal{S}\left( \mathbf{O}_{1:T}\right) $ comprise the first
four moments and first lag autocorrelation statistic of each $\widehat{%
\sigma }_{t}^{\left( i,j\right) }$ term from the expansion, as well as the
average standardized distance between the simulated and observed $\widehat{%
\sigma }_{t}^{\left( i,j\right) }$. 

\section*{Appendix C: Further Simulation Results}

\setcounter{table}{0} \renewcommand{\thetable}{C%
\arabic{table}} \setcounter{figure}{0} \renewcommand{\thefigure}{C%
\arabic{figure}}

\subsubsection*{ C.1 Comparison with Conventional ABC}

Table \ref{tab:HestionOption} reports the
empirical coverage of the 80\%, 90\% and 95\% prediction intervals produced
using \eqref{eq:approxpred}, based on the modular ABI algorithm
(Panel A), and the conventional ABC accept-reject algorithm applied \textit{without} cutting feedback (Panel
B) respectively. Conventional ABC is implemented using the
misspecified simulator as described in Section \ref{sec:mc_noise}, and with the tuning
parameter $\epsilon $ in Algorithm \ref{alg:abc} chosen such that we retain
the 5\% of the draws deemed to produce the closest summary statistics to the
observed statistics. With coverages that are closest to the nominal coverage
preferred, it is clear that the modular ABI predictives generate coverage
statistics that are uniformly closer to the nominal level than are the
corresponding statistics for the standard ABC algorithm. In particular, the
conventional ABC accept-reject algorithm produces predictive intervals that have
empirical coverages that are too large, reflecting the larger degree of
uncertainty in the posteriors for the model parameters. This excess coverage
becomes more pronounced as the option maturity increases, with the
structural parameters playing a more important role in the risk-neutral
expectation that forms the theoretical price of longer-term options compared
to shorter-term options. These results confirm the superior performance of
our modular ABI approach over a conventional ABC approach. 

\spacingset	{1.0} 
\begin{table}[tbph]
\caption{Out-of-sample prediction interval coverage for option prices. Panel
A reports empirical coverages from the modular ABI approach; Panel B reports
empirical coverages from the conventional ABC accept-reject method.}
\label{tab:HestionOption}
\centering \medskip 
{\footnotesize \ 
\begin{tabular}{l|rrr|rrr}
& \multicolumn{3}{c|}{Panel A: Modular ABI} & \multicolumn{3}{c}{Panel B:
Conventional ABC} \\ 
Maturity & 80\% & 90\% & 95\% & 80\% & 90\% & 95\% \\ \hline
{5} & 80.7\% & 91.4\% & 95.0\% & 81.1\% & 91.7\% & 95.1\% \\ 
{10} & 80.8\% & 91.2\% & 94.9\% & 81.1\% & 91.8\% & 95.0\% \\ 
{15} & 80.6\% & 91.3\% & 94.9\% & 81.6\% & 92.0\% & 95.2\% \\ 
{20} & 80.7\% & 91.4\% & 95.0\% & 82.7\% & 92.3\% & 95.8\% \\ 
{25} & 80.8\% & 91.6\% & 95.2\% & 84.1\% & 92.9\% & 96.2\% \\ 
{30} & 81.2\% & 91.7\% & 95.3\% & 85.3\% & 93.7\% & 96.6\% \\ 
{60} & 83.9\% & 92.8\% & 96.5\% & 91.0\% & 96.3\% & 98.4\% \\ \hline
Overall & 81.2\% & 91.6\% & 95.2\% & 83.8\% & 93.0\% & 96.1\% \\ \hline\hline
\end{tabular}%
}
\end{table}

\spacingset{1.8}

\subsubsection*{C.2 The Impact of the Misspecified Simulator}

Using the simulation experiment setting described in
Section \ref{sec:mc_noise}, we document the impact of using the misspecified
simulator in the construction of the soft cut posterior. Panel A of Table %
\ref{tab:misspecified} documents the prediction coverage using the
``correct'' simulator, where the cut posterior is obtained by
simulation from \eqref{eq:option_noise} and \eqref{eq:hf_noise}. Panel B
documents the results from the cut posterior obtained from the
``misspecified'' simulator, as described in Section \ref{sec:simulator}.
While we observe some minor differences in the cut posteriors, use of the misspecified
simulator has very little impact on the approximate predictive
distribution of the option prices, as can be seen in the predictive
coverages reported in Table \ref{tab:misspecified} below. 

\spacingset{1.0} 
% Table generated by Excel2LaTeX from sheet '3kgrid'
\begin{table}[tbph]
\caption{Out-of-sample prediction interval coverage for option prices when
data is contaminated with measurement error and microstructure noise. Panel
A reports empirical coverages from the modular ABI approach using
the ``correct'' simulator as prescribed by
the DGP; Panel B reports empirical coverages from the modular ABI
approach using the proposed ``misspecified'' simulator.}
\label{tab:misspecified}  
\centering \medskip %
 \spacingset	{1.0}{\footnotesize 
\begin{tabular}{l|rrr|rrr}
\multicolumn{1}{c}{} & \multicolumn{3}{c}{} & \multicolumn{3}{c}{} \\ 
& \multicolumn{3}{c|}{Panel A: ``Correct'' simulator} & \multicolumn{3}{c}{Panel B: ``Misspecified'' simulator} \\ 
Maturity (Days) & 80\% & 90\% & 95\% & 80\% & 90\% & 95\% \\ \hline
{5} & 80.4\% & 89.9\% & 95.1\% & 80.6\% & 90.4\% & 95.6\% \\ 
{10} & 80.3\% & 90.3\% & 95.1\% & 80.4\% & 90.8\% & 95.5\% \\ 
{15} & 80.7\% & 90.6\% & 95.3\% & 80.2\% & 91.0\% & 95.4\% \\ 
{20} & 81.1\% & 90.9\% & 95.6\% & 80.4\% & 91.0\% & 95.5\% \\ 
{25} & 81.5\% & 91.2\% & 95.7\% & 80.5\% & 91.1\% & 95.6\% \\ 
{30} & 81.9\% & 91.7\% & 95.9\% & 81.1\% & 91.4\% & 95.8\% \\ 
{60} & 83.6\% & 93.3\% & 96.6\% & 83.7\% & 92.5\% & 96.8\% \\ \hline
Overall & 81.4\% & 91.1\% & 95.6\% & 81.0\% & 91.2\% & 95.7\% \\ \hline\hline
\end{tabular}%
}  
\end{table}

\spacingset{1.8}

\spacingset	{1.0}

\section*{Appendix D: Robustness to Gaussian Mixture Model Approximation Methods} \label{app:GMMfit}

\spacingset{1.8}%
\setcounter{table}{0} \renewcommand{\thetable}{D\arabic{table}} %
\setcounter{figure}{0} \renewcommand{\thefigure}{D\arabic{figure}} 

We employ an alternative algorithm to estimate the
parameters of the Gaussian mixture model to assess the sensitivity of the
resulting soft cut posterior to the method used to fit the Gaussian
mixture. We compare the optimal cut posterior 
documented in the paper, and as computed using the MATLAB
expectation-maximization (EM) algorithm, with the posterior
produced via the telescope sampler of \cite%
{fruhwirth2021generalized}, using the `telescope' package in R. The marginal
posterior distributions obtained using the two versions of the 
fitted Gaussian mixture model are displayed in Figure \ref%
{fig:compare}. As in the optimal cut structure assessment, we also 
produced the posterior mass around the true parameter values, with
the Gaussian mixture model fitted using the telescope sampler
returning slightly lower relative posterior mass (0.905) 
than the optimal cutting structure based on the EM algorithm (0.944). This is unlikely to have any significant impact on predictive 
performance, as verified by the documented predictive coverages in Table \ref{tab:optcov}.

\begin{figure}[tbh]
\centering
\includegraphics[width=1\textwidth]{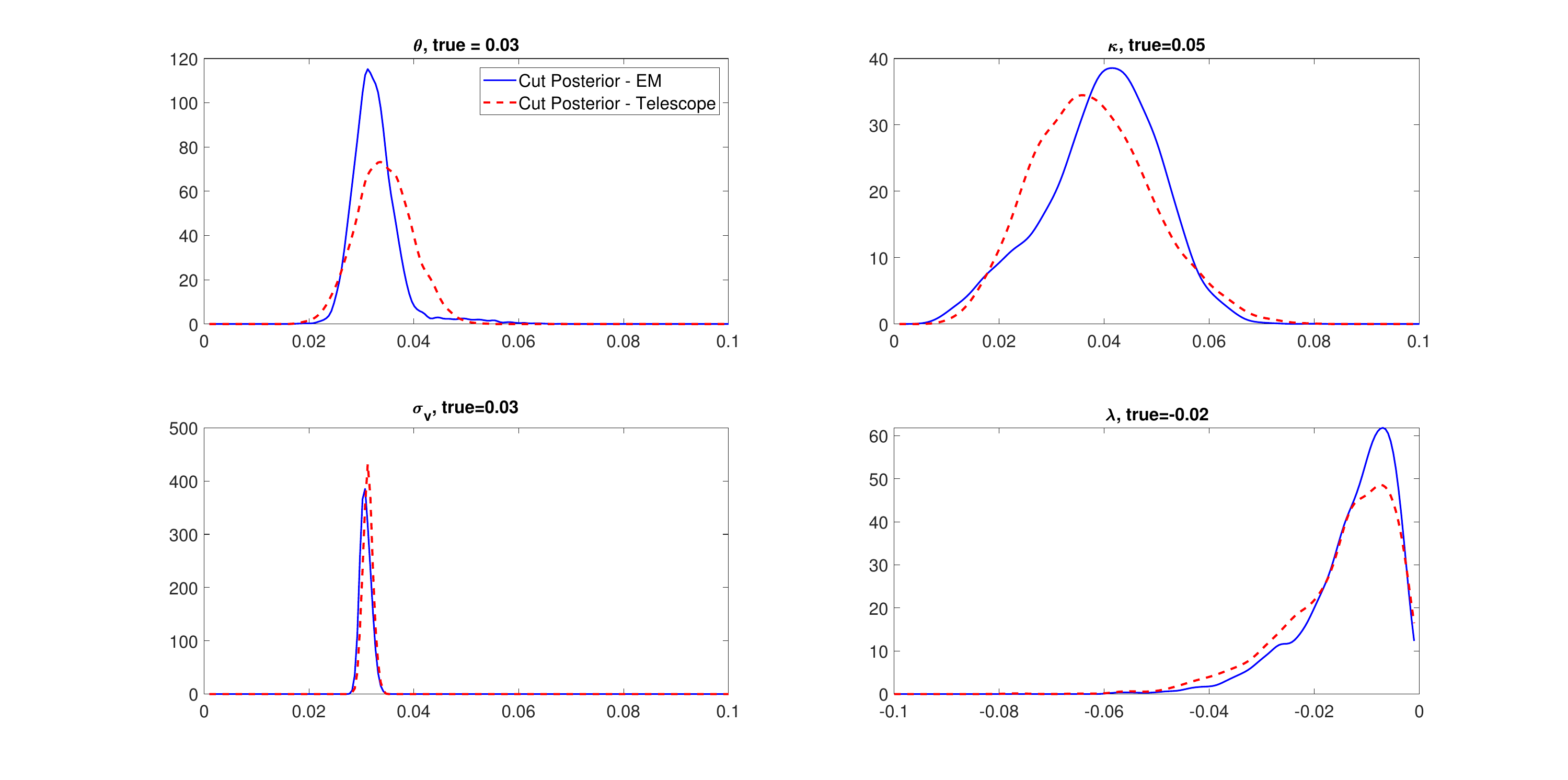}  
\caption{Comparison of cut posteriors obtained from the Gaussian mixture
model fitted using MATLAB's Expectation-Maximization algorithm and the posterior mode implied by the telescope sampler.}
\label{fig:compare}
\end{figure}
 \spacingset	{1.0} 
\begin{table}[tbph]
\caption{Out-of-sample prediction interval coverage from the optimal cut
structure. Panel A reports empirical coverages from the Gaussian mixture
model estimated using the Telescope sampler; Panel B reports those from the
EM algorithm.}
\label{tab:optcov}  
\centering \medskip {\footnotesize 
\begin{tabular}{l|rrr|rrr}
& \multicolumn{3}{c|}{Panel A: Telescope} & \multicolumn{3}{c}{Panel B: EM}
\\ 
Maturity (Days) & 80\% & 90\% & 95\% & 80\% & 90\% & 95\% \\ \hline
{5} & 80.4\% & 90.6\% & 94.6\% & 80.7\% & 90.9\% & 94.5\% \\ 
{10} & 81.1\% & 91.1\% & 95.0\% & 81.1\% & 91.3\% & 94.9\% \\ 
{15} & 81.6\% & 91.3\% & 95.3\% & 81.5\% & 91.6\% & 95.3\% \\ 
{20} & 82.0\% & 91.6\% & 95.6\% & 81.7\% & 92.0\% & 95.5\% \\ 
{25} & 82.5\% & 92.0\% & 95.7\% & 82.0\% & 92.4\% & 95.7\% \\ 
{30} & 82.9\% & 92.4\% & 95.8\% & 82.3\% & 92.6\% & 95.9\% \\ 
{60} & 85.2\% & 94.1\% & 96.7\% & 84.5\% & 93.6\% & 96.4\% \\ \hline
Overall & 82.2\% & 91.9\% & 95.5\% & 82.0\% & 92.1\% & 95.5\% \\ \hline\hline
\end{tabular}%
}
\end{table}

\spacingset	{1.8}

\section*{Appendix E: Further Empirical Results}

\setcounter{table}{0} \renewcommand{\thetable}{E%
\arabic{table}} \setcounter{figure}{0} \renewcommand{\thefigure}{E%
\arabic{figure}}

\subsubsection*{E.1 Option Contracts Summary}

Figure \ref{fig:optionContracts} depicts the number of
actively traded call option contracts over the sample period employed for
the empirical analysis. The number of actively traded contracts has
increased substantially over the past decade.

% Figure \ref{fig:fixedWindow} summarizes the evolution
% of the (cut) posterior median for each parameter, over the rolling-window
% samples. Reiterating that the Heston pricing model is almost certainly
% misspecified, we observe distinct changes in the structural parameters of
% the model over time. The unconditional variance, $\theta $ and the
% volatility of volatility $\sigma _{v}$, both increase with the level of
% market volatility, peaking during the midst of the Global Financial Crisis
% between 2008 and 2009, and more recently during the COVID-19 pandemic. We
% also note that the persistence parameter $\kappa $ is relatively high
% compared to previous studies, most likely due to the lack of incorporation
% of jumps in the pricing process. The risk premium parameter $\lambda $ also
% changes overtime, indicating the changing compensation for random volatility
% required by the option market.

%It is important to stress here that, with the Heston
%model being misspecified,\ any particular posterior estimates should be
%taken with a grain of salt; the time-variation and extreme changes in the
%features of the posterior distributions signalling the inadequacies of the
%Heston model. However, \textbf{as documented in the main paper, our fast} posterior %updating scheme \textbf{yields} \textbf{predictive distributions for option prices %that adequately reflect the day-to-day movements in the financial markets, and %produce accurate predictive coverages as a consequence, for short-maturity options in %particular.}

\spacingset	{1.0} 
\begin{figure}[tbh]
\centering
\includegraphics[width=0.8\textwidth]{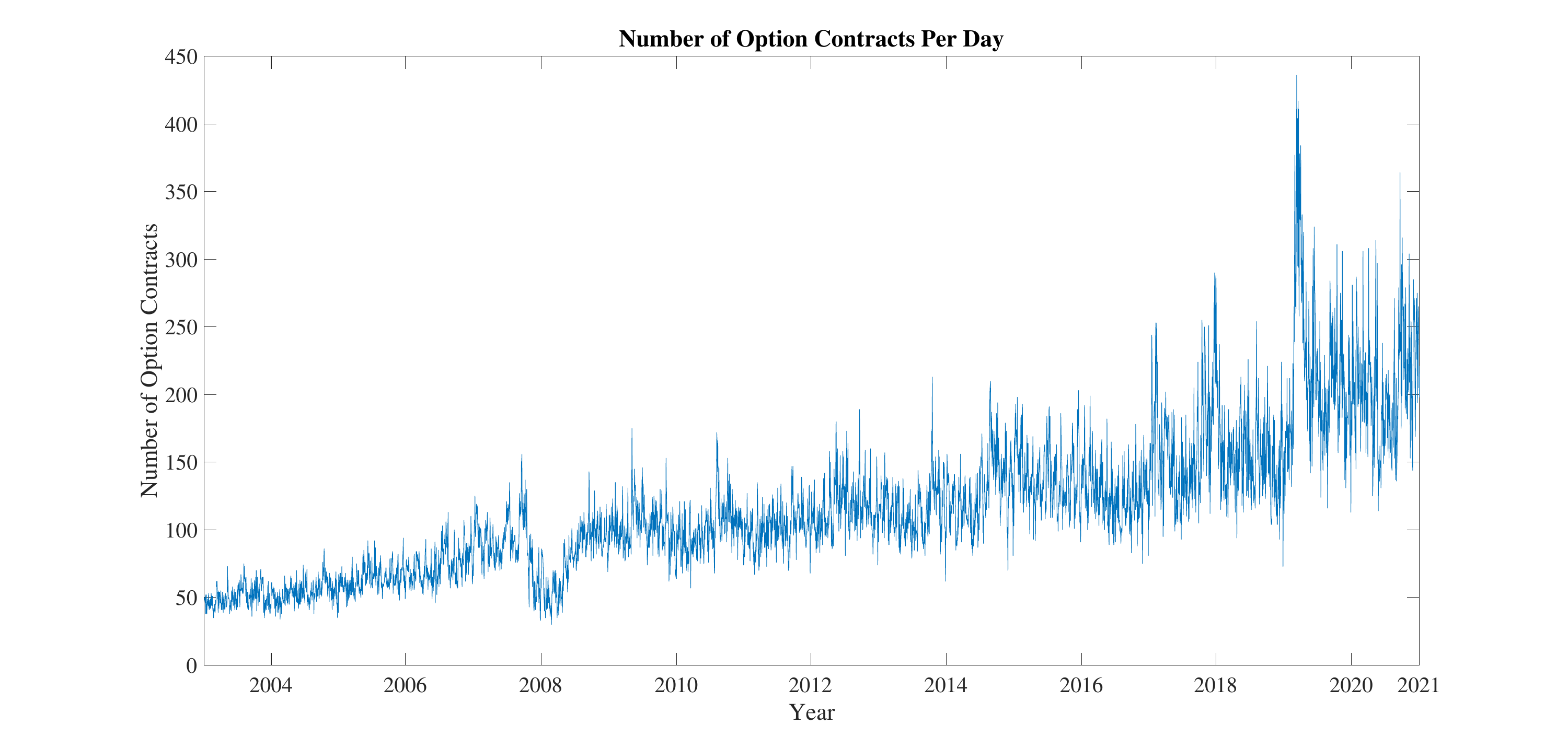}  
\caption{Number of active call option contracts per trading day that have
trade volume equals to or larger than 10 traded contracts per day and strike prices $\pm
20\%$ of the S\&P500 index value.}
\label{fig:optionContracts}
\end{figure}

% \begin{figure}[tbh]
% \centering
% \includegraphics[width=0.9\textwidth]{empEstOpt.eps}  
% \caption{Posterior median of the Heston model parameters, using the fixed
% rolling-window estimation scheme. The risk-free interest rate over the
% estimation period is taken to be the average Federal Fund rate over the
% estimation window.}
% \label{fig:fixedWindow}
% \end{figure}
 \spacingset	{1.8}

\subsubsection*{E.2 Additional Predictive Results}

Table \ref{tab:pred_opt1} documents more detailed
coverage results by maturity and Table \ref{tab:pred_byk} documents further
predictive coverage results for the 90\% prediction interval, segmented by
option moneyness $|k|$.

\begin{table}[htbp]
\caption{Empirical coverage of the predictive distribution of option prices
by maturity.}
\label{tab:pred_opt1}
\centering \medskip %
 \spacingset	{1.0}{\footnotesize 
\begin{tabular}{rrrrr}
\multicolumn{1}{l}{Maturity} & {95\%} & {90\%} & {80\%} & \multicolumn{1}{l}{
\# of contracts} \\ \hline
1 day & 93.6\% & 89.9\% & 80.0\% & 3681 \\ 
2 days & 90.5\% & 86.4\% & 77.0\% & 6402 \\ 
3 days & 89.7\% & 84.6\% & 73.4\% & 6998 \\ 
4 days & 91.3\% & 84.6\% & 74.0\% & 6946 \\ 
5 days & 91.4\% & 85.8\% & 76.9\% & 3144 \\ \hline
1 week & 90.5\% & 84.8\% & 74.2\% & 39014 \\ 
2 weeks & 86.5\% & 79.0\% & 67.5\% & 38458 \\ 
3 weeks & 83.8\% & 75.9\% & 64.3\% & 38817 \\ 
4 weeks & 77.0\% & 68.0\% & 55.5\% & 40841 \\ \hline
1 month & 82.9\% & 75.2\% & 63.6\% & 176808 \\ 
2 months & 69.5\% & 60.4\% & 49.0\% & 152035 \\ 
3 months & 65.6\% & 56.9\% & 45.8\% & 86971 \\ 
4 months & 68.8\% & 60.2\% & 48.8\% & 43386 \\ 
5 months & 67.4\% & 58.4\% & 46.8\% & 21118 \\ 
6 months & 61.4\% & 52.9\% & 42.2\% & 16569 \\ \hline\hline
Overall & 73.2\% & 64.7\% & 53.3\% & 496887 \\ \hline\hline
\end{tabular}%
 \spacingset	{1.8}}  
\end{table}

 \spacingset	{1.0} 
\begin{table}[tbph]
\caption{Empirical coverage of the 90\% prediction interval of option
prices. Results are reported by option maturity up to one month, and
segmented by the relative deviation between the strike price \textbf{} the spot price, $%
|k|$.}
\label{tab:pred_byk}
\centering \medskip %
 \spacingset	{1.0} {\footnotesize \ 
\begin{tabular}{r|cccc}
\multicolumn{1}{l}{} &  &  &  &  \\ \hline\hline
& \multicolumn{4}{c}{$|k|$} \\ 
\multicolumn{1}{l|}{Maturity} & \multicolumn{1}{l}{0 - 5\%} & 
\multicolumn{1}{l}{\ 5 - 10\%} & \multicolumn{1}{l}{\ 10 - 15\%} & 
\multicolumn{1}{l}{\ 15 - 20\%} \\ \hline\hline
1 day & 91.4\% & 85.3\% & 80.7\% & 73.7\% \\ 
2 days & 87.1\% & 83.0\% & 86.2\% & 83.3\% \\ 
3 days & 87.7\% & 74.6\% & 64.8\% & 72.7\% \\ 
4 days & 86.9\% & 77.6\% & 73.2\% & 69.5\% \\ 
5 days & 88.8\% & 81.4\% & 64.5\% & 52.2\% \\ 
2 weeks & 79.0\% & 80.7\% & 74.0\% & 64.9\% \\ 
3 weeks & 75.8\% & 75.1\% & 81.6\% & 73.8\% \\ 
1 month & 76.9\% & 70.1\% & 76.1\% & 73.8\% \\ \hline\hline
\end{tabular}
}  
\end{table}

\end{document}